%% file: main.tex
\documentclass[twocolumn]{aastex62}

\usepackage{psfig,enumerate}
\usepackage{url}
\usepackage{verbatim}
\usepackage{color}
\usepackage{atbegshi}
\usepackage{bm}

\input{shorthand.tex}

\begin{document}

\title{The Second Data Release of the Survey of the MAgellanic Stellar History (SMASH)}

\shorttitle{SMASH DR2}
\shortauthors{Nidever et al.}

\input{authors.tex}

\begin{abstract}
The Large and Small Magellanic Clouds (LMC and SMC) are the largest satellite galaxies of the Milky Way and close enough to allow for a detailed exploration of their structure and formation history. The Survey of the MAgellanic Stellar History (SMASH) is a community Dark Energy Camera (DECam) survey of the Magellanic Clouds
using $\sim$50 nights to sample over $\sim$2400 deg$^2$ centered on the Clouds at $\sim$20\% filling factor (but with contiguous coverage in the central regions) and to depths of $\sim$24th mag in $ugriz$.  The primary goals of SMASH are to map out the extended stellar peripheries of the Clouds and uncover their complicated interaction and accretion history as well as to derive spatially-resolved star formation histories of the central regions and create a ``movie'' of their past star formation.  Here we announce the second SMASH public data release (DR2), which contains all 197 fully-calibrated DECam fields including the main body fields in the central regions.  The DR2 data are available through the Astro Data Lab hosted by the NSF's National Optical-Infrared Astronomy Research Laboratory.  We highlight three science cases that make use of the SMASH DR2 data and will be published in the future:  (1) preliminary star formation histories of the LMC; (2) the search for Magellanic star clusters using citizen scientists; and, (3) photometric metallicities of Magellanic Cloud stars using the DECam $u$-band.

\end{abstract}

\keywords{dwarf galaxy: individual: Large Magellanic Cloud, Small Magellanic Cloud --- Local Group --- Magellanic Clouds --- surveys}

\section{Introduction}
\label{sec:intro}

The relative proximity ($\sim$ 55 kpc) of the Large and Small Magellanic Clouds (LMC and SMC)---the largest satellite galaxies of the Milky Way (MW)---presents opportunities to perform detailed resolved studies on their stellar populations and to uncover their structure, evolution, and past interactions. Systems like the Magellanic Clouds (MCs) are rare in the local galactic neighborhood because they are in close proximity to their host, are a binary pair \citep {besla2018}, and have vigorous ongoing star formation \citep[e.g.,][]{Harris2004,harris_zaritsky2009,Weisz2013,meschin2014, rubele2018}. Their active star formation is partially due to their large masses (M$_{\text{LMC}} = 1.38 \pm 0.24 \times 10^{11}$ $\text{M}_\odot$, \citealt{Erkal2019} and M$_{\text{SMC}}$ = 2.1 $\times$ 10$^{10}$ M$_\odot$, \citealt{Besla2012}), which allows them to retain their gas longer, but also due to the fact they are on their first infall into the MW potential \citep{Besla2007,Kallivayalil2006,Kallivayalil2013} and have therefore gone largely unscathed from the inhospitable environment close to large galaxies.

The MCs subtend many hundreds of square degrees in the sky, which makes them challenging to study comprehensively.  Apart from the photographic plate \citep{devaucouleurs55, devaucouleurs55b, Irwin1991, GardinerHatzidimitriou1992} and wide-field CCD works \citep{BothunThompson1988}, most early studies of field MC stars focused on a number of small regions in them, providing the very first CMDs of their bright stars \citep[e.g.][]{Westerlund1970, FrogelBlanco1983, Stryker1984field, Hardy1984LMC, Hodge1987LMC_field, Bertelli1992}. More recently, large ground-based CCD observing campaigns mapped the inner regions of the MCs (Optical Gravitational Lensing Experiment, OGLE, \citealt{Udalski1992}; Magellanic Clouds Photometric Survey, MCPS, \citealt{Zaritsky2002}; SuperMACHO, \citealt{Rest2005}). 
Deep ground-based observations were able to derive ages for MC and Galactic clusters \citep[e.g.,][]{Brocato1996}.  Soon thereafter,
the Hubble Space Telescope ($HST$) provided very deep CMDs reaching well below the oldest main sequence turnoffs \citep[e.g.,][]{Elson1997, Olsen1999, Holtzman1999, Smecker-Hane2002, Cignoni2013}, albeit for very small fields of view in the crowded inner regions. Soon after it was shown that similar quality data could be obtained from the ground for less crowded areas (see \citealt{Monteagudo2018} for ground based CMDs reaching the oldest Main Sequence Turn-Off (oMSTO) for the inner disk and bar of the LMC). In this way, the outer disk and periphery of the LMC was explored with island fields by \citet{Gallart2004, Gallart2008}, \citet{Majewski2009} and \citet{Saha2010}, while the SMC was targeted by \citet{Noel2007} and \citet{Nidever2011}. \citet{Saha2010} and \citet{Majewski2009} found extended LMC stellar populations out to $\sim$16--20\dgr while \citet{Nidever2011} found red giant branch stars out to a radius of 11\dgr in the SMC. In the last decade, the advent of large field-of-view, multi-CCD detectors has greatly enhanced the ability to map the MCs.  The VISTA survey of the Magellanic Clouds system \citep[VMC;][]{Cioni2011} is mapping the main bodies of the MCs in near-infrared bands, while the STEP \citep[The SMC in Time: Evolution of a Prototype interacting late-type dwarf galaxy][]{Ripepi2014} and YMCA \citep[Yes, Magellanic Clouds Again][]{Gatto2020} surveys are using the VLT Survey Telescope (VST) to map the Magellanic periphery in optical bands.
The construction of the Dark Energy Camera \citep[DECam;][]{Flaugher2015} on the Blanco 4$-$m telescope at Cerro Tololo Inter-American Observatory (CTIO) has enabled deep and efficient imaging of the whole southern sky in the last seven years.  The Dark Energy Survey \citep[DES;][]{DES} mapped $\sim$5,000 square degrees including part of the MCs, using data obtained with DECam. There are two ongoing surveys that also use DECam to study areas around the MCs: (1) the Magellanic Satellites Survey \citep[MagLites;][]{torrealba2018}, which will map $\sim$1,200 deg$^2$ near the south celestial pole and (2) the DECam Local Volume Exploration \citep[DELVE;][]{mau2020} survey that has a MC component that will map $\sim$1,000 deg$^2$ of the Magellanic periphery. Both MagLites and DELVE have goals to search for ultra--faint dwarf galaxy satellites near the LMC and SMC.

\begin{figure*}[t]
\includegraphics[width=1.0\hsize,angle=0]{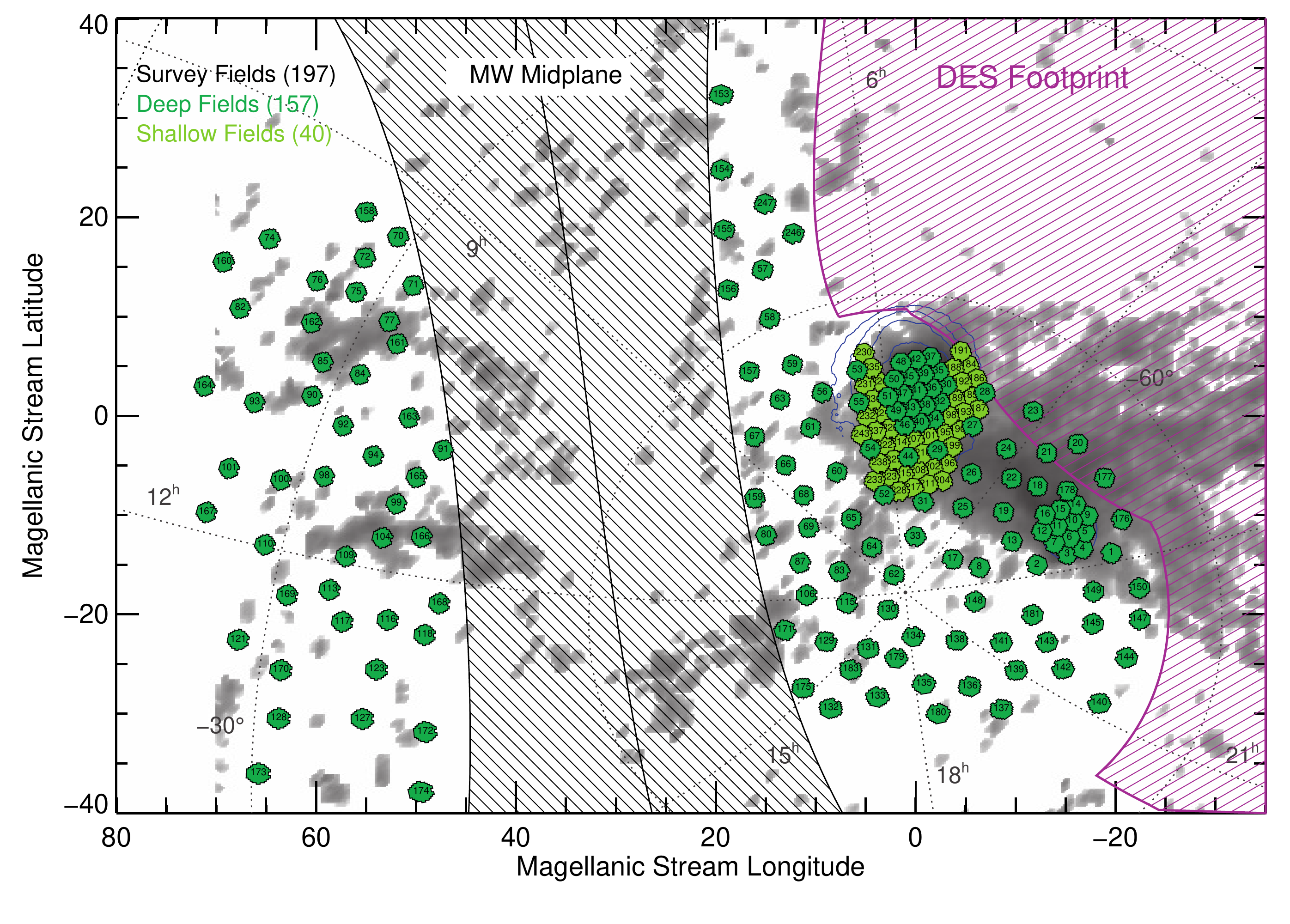}
\caption{Map of the SMASH survey.  The \citet{Nidever2010} observed \hi column density of the Magellanic system is shown in the background grayscale.  The SMASH fields are shown as filled green hexagons (dark green: deep, light green: shallow).  The Milky Way mid-plane extending to $\pm$15\dgr in latitude is indicated by the black shaded region and was avoided by SMASH.
The purple shaded region indicates the DES footprint.
}
\label{fig_dr2map}
\end{figure*}

The Survey of the MAgellanic Stellar History \citep[SMASH;][]{Nidever2017} is a large, community NOAO\footnote{NOAO is now wholly encompassed in the NSF’s National Optical-Infrared Research Laboratory under the programs Mid-scale Observatories (MSO) and Community Science and Data Center. The Kitt Peak National Observatory and Cerro Tololo Inter-American Observatory are both part of MSO.} survey of the MCs using $\sim$50 nights of DECam time. SMASH maps over 480 square degrees of the sky, contained within $\sim$2400 square degrees of the Magellanic System (see Fig.\ \ref{fig_dr2map}) complementary to the DES coverage.  The main goals of SMASH are to (a) search for low surface-brightness structure in the periphery of the MCs, (b) search for the stellar component of the Magellanic Stream and Leading Arm, and (c) derive precise, spatially-resolved star formation histories (SFHs) to ancient ages.  \citet{Nidever2017} have laid out the survey details (strategy, observations, data reduction and calibration) and presented the first data release (DR1) of 61 SMASH fields as well as some science examples.

Besides the main goals outlined above, SMASH data have produced many science results. We serendipitously discovered the Hydra II dwarf galaxy \citep{Martin:2015} and used follow-up observations to discover one RR Lyrae star in this satellite \citep{Vivas:2016}. A more extensive search for stellar overdensities detected SMASH 1, which appears to be a tidally disrupting globular cluster in the periphery of the LMC \citep{Martin:2016}. \citet{Choi:2018b} used SMASH data to map the 2D dust reddening and 3D structure of the LMC, confirming previous findings of a stellar warp in the inner disk and discovering a major stellar warp in the outer disk. SMASH also demonstrated the existence of a ring-like stellar overdensity in the LMC, which was likely created by recent, close tidal interactions with the SMC \citep{Choi:2018a}. \citet{Nidever2019a} found evidence of old LMC MSTO stars out to roughly 18.5 kpc from the LMC center, while \citet{Massana2020} detect the disturbed stellar structures in the periphery of the SMC out to $\sim$8 kpc.  \citet{Martinez-Delgado:2019} investigated a shell of stars located $1.9 \degr$ from the SMC center, and found that it is composed of young stars, suggesting a recent star formation event triggered by a direct collision between the Clouds.
\citet{Bell2019,Bell2020} used a combination of SMASH and VMC data to map the total intrinsic extinction through the SMC using spectral energy distributions of background galaxies. 

In this paper we describe the second public data release from the SMASH survey, which includes---for the first time---DECam data in the central regions of the Magellanic Clouds, and we present preliminary analyses of the science cases that have so far been conducted using SMASH DR2.  
The layout of this paper is as follows.  Section \ref{sec:observing} briefly describes our final observations while Section \ref{sec:reduction} outlines some of the modifications of the data processing compared to DR1.  Section \ref{sec:performance} describes the data release files and the achieved performance while Section \ref{sec:dr2} reviews the capabilities available in the Astro Data Lab.  Finally, Section \ref{sec:examples} presents three science examples using SMASH DR2 data.

\section{Observations}
\label{sec:observing}

Some additional observations were obtained after the release of SMASH DR1 \citep{Nidever2017}.  Most of these were using Director's Discretionary (DD) time or Engineering time to obtain short exposures during photometric conditions to calibrate previous data that were obtained in non-photometric conditions. The full list of SMASH observations are summarized in Table 2. The median seeing in the observations is 1.22\arcsec\ ($u$), 1.13\arcsec\ ($g$), 1.01\arcsec\ ($r$), 0.95\arcsec\ ($i$), and 0.90\arcsec\ ($z$) with a scatter of roughly 0.25\arcsec.


\section{Data Processing and Calibration}
\label{sec:reduction}

The SMASH data reduction is described in detail in \citet{Nidever2017}, so we only give a short overview here.  SMASH makes use of the DECam calibrated data produced by the NOAO Community Pipeline \citep[CP;][]{Valdes2014}\footnote{\url{http://www.noao.edu/noao/staff/fvaldes/CPDocPrelim/PL201\_3.html}} which performs the instrument signature removal (e.g., bias subtraction, flat fielding, etc.).  The PHOTRED\footnote{\url{https://github.com/dnidever/PHOTRED}} pipeline \citep{PHOTRED} is then used to perform PSF photometry on each individual exposure first and forced-photometry on all overlapping exposures later with DAOPHOT/ALLFRAME \citep{Stetson1994}.  Finally, custom calibration software (SMASHRED\footnote{\url{https://github.com/dnidever/SMASHRED}}; \citealt{SMASHRED}) is used to perform an ubercal-like calibration \citep{Padmanabhan2008} of exposures in a field to combine measurements and to calibrate the photometry.

We obtained multiple deep ($\sim$300s) and shallow (60s) exposures in every filter and a majority of our fields (with the exception of the ``shallow'' fields in the LMC).  The deep exposures allow us to reach depths of 24--25th mag, while the shallow exposures fill in the brighter magnitudes (to $\sim$16th mag, although this is color dependent; see Fig.\ \ref{fig_cmds}). The completeness is reduced for the brightest stars, such as the tip of the red giant branch, the asymptotic giant branch, and the youngest main-sequence stars.  Therefore, we suggest some caution when studying these populations in the SMASH catalogs.

A substantial fraction of the SMASH fields are ``island'' fields that do not overlap other SMASH fields (see Fig.\ \ref{fig_dr2map}).  Therefore, during the standard calibration and combination steps the exposures of only one field were dealt with at a time.
However, SMASH has contiguous coverage of the central regions (or ``main bodies'') of the MCs and the fields there do overlap. Providing catalogues on a field-by-field basis would produce a large number of duplicate objects in the main bodies.  To remedy this issue, an additional step was taken to combine the data uniformly in the MC main bodies.  The data were combined and calibrated in a similar manner to the regular fields but using HEALPix \citep{Gorski2005} instead of fields.  An {\tt nside=64} was used, corresponding to an area of 0.89 square degrees and resulting in 321 total HEALPix with 257/64 for LMC/SMC.
The field-level ubercal zeropoint corrections were used for calibrating the data in HEALPix regions, because in some cases (HEALPix at the edges of the contiguous regions) the number of overlapping chips was too small to reliably perform the ubercal calibration.
In the main body regions, both HEALPix and field-level catalogs are provided in the flat FITS files (see Section \ref{sec:dr2}), but only the HEALPix catalogs are loaded in the {\tt object} table of the database (see Section \ref{subsec:catalogdescription}) in order to remove duplicates.

\begin{figure*}[t]
\includegraphics[width=1.0\hsize,angle=0]{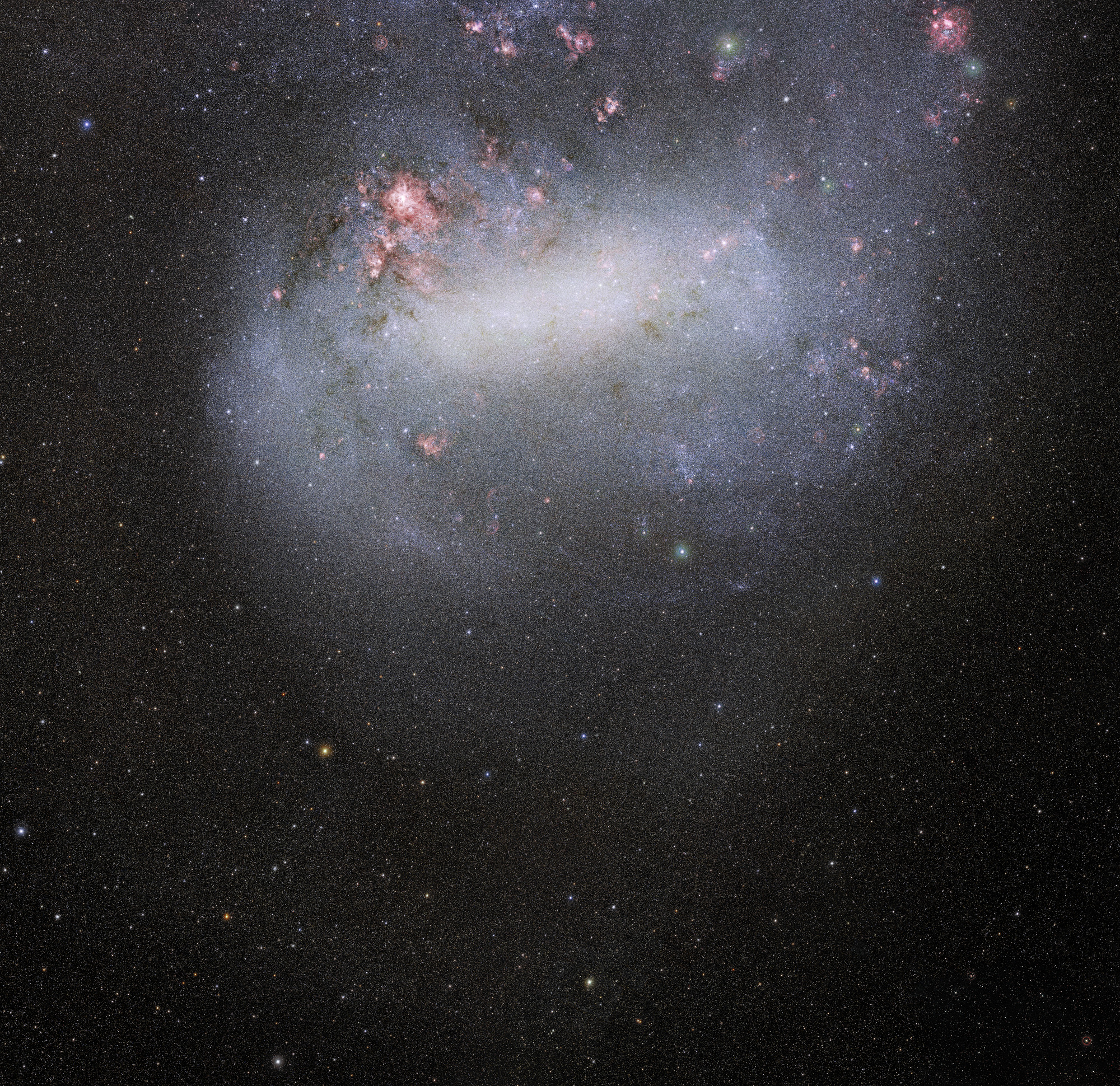}
\caption{A three-color image of the SMASH fields in the main body of the Large Magellanic Cloud. Credit: CTIO/NOIRLab/NSF/AURA.}
\label{fig_lmc3color}
\end{figure*}

\begin{figure*}[t]
\includegraphics[width=1.0\hsize,angle=0]{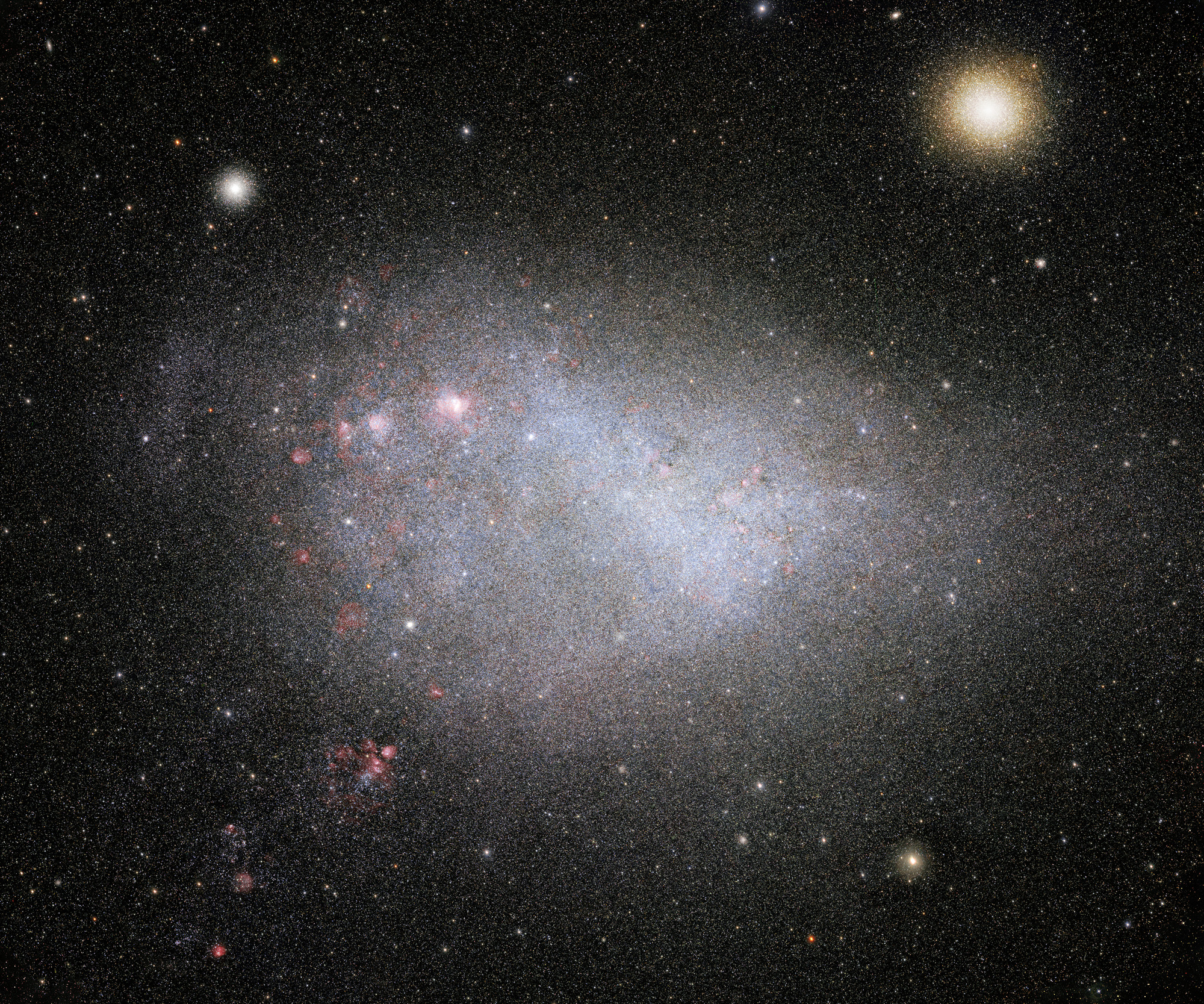}
\caption{A three-color image of the SMASH fields in the main body of the Small Magellanic Cloud. Credit: CTIO/NOIRLab/NSF/AURA.}
\label{fig_smc3color}
\end{figure*}

\begin{figure*}
\begin{center}
$\begin{array}{cc}
\includegraphics[width=0.48\hsize,angle=0]{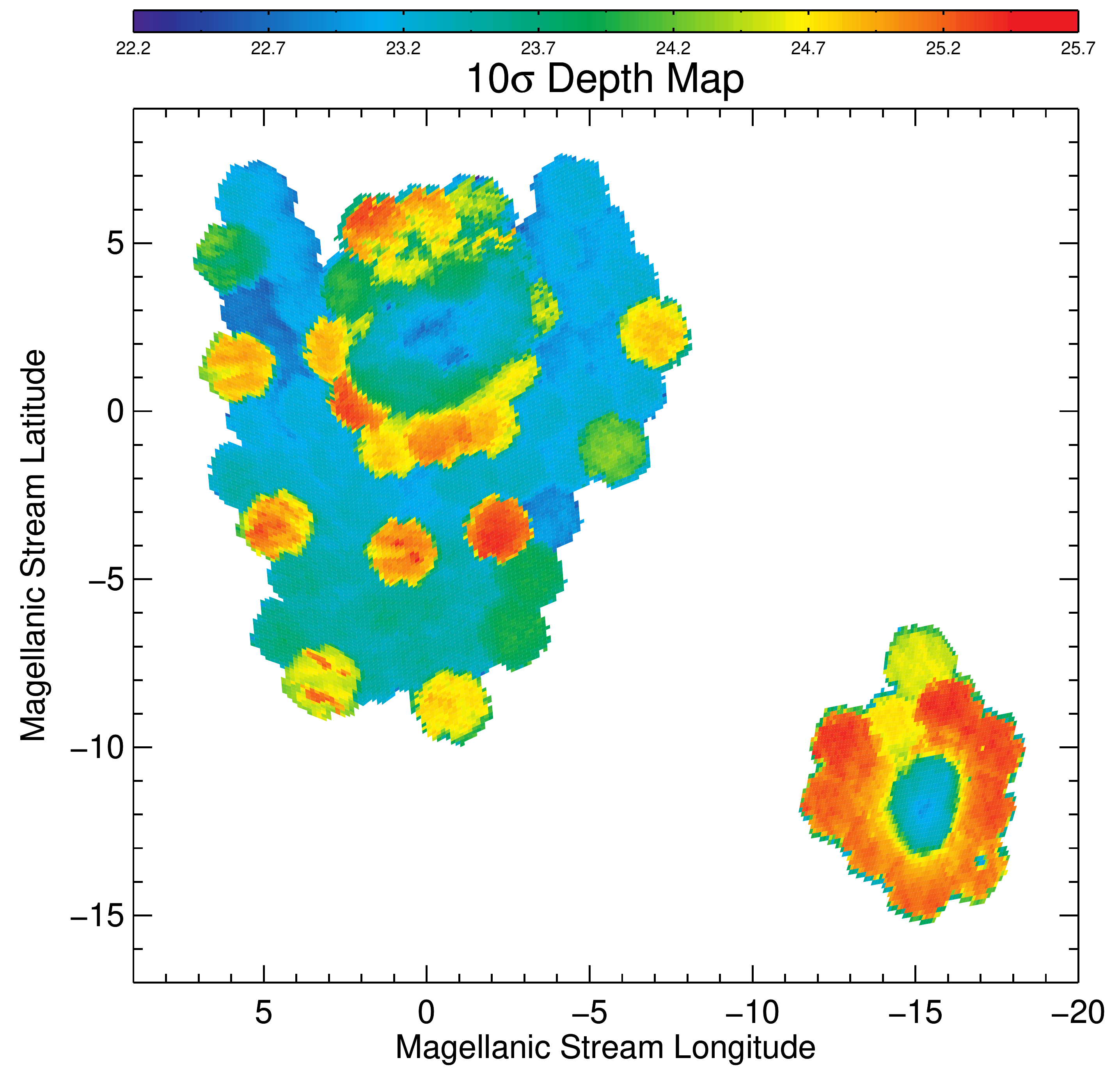}
\includegraphics[width=0.48\hsize,angle=0]{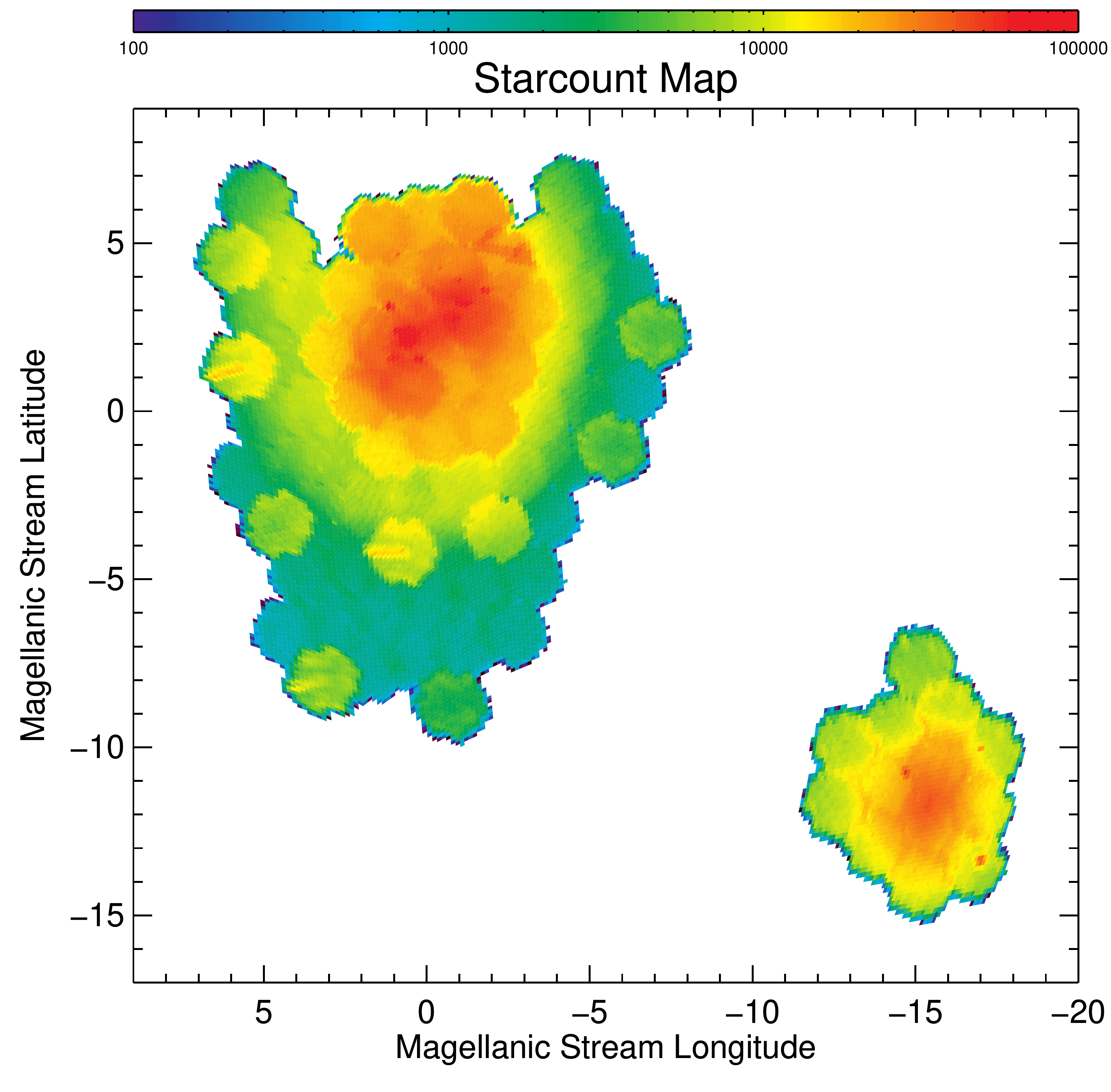}
\end{array}$
\end{center}
\caption{(Left) The SMASH DR2 $g$-band 10$\sigma$ depth in the contiguous regions in the main bodies of the Magellanic Clouds using nside=512 HEALPix ($\sim$7\arcmin~across). (Right) Number of objects in each HEALPix over the same region.  There is reduced depth in the very crowded centers of the MCs.  The 40 shallow LMC fields at somewhat larger radii can also be seen by their reduced depth.}
\label{fig_depth}
\end{figure*}

\section{Description and Achieved Performance of the Final Catalogs}
\label{sec:performance}

The SMASH DR2 dataset includes 5,982 DECam exposures with 359,393 separate chip files producing 4,156,198,451 independent source measurements of 360,363,299 unique objects.

\subsection{Final Catalog Files}
\label{subsec:catalogdescription}

The final catalogs consist of eight gzip-compressed binary FITS files per field and HEALPix:
\begin{enumerate}
\item {\tt FIELD/HEALPix\_exposures.fits.gz} -- Information about each exposure.
\item {\tt FIELD/HEALPix\_chips.fits.gz} --  Information about each chip.
\item {\tt FIELD/HEALPix\_allsrc.fits.gz} -- All of the individual source measurements for this field.
\item {\tt FIELD/HEALPix\_allobj.fits.gz} -- Average values for each unique object.
\item {\tt FIELD/HEALPix\_allobj\_deep.fits.gz} -- Average values for each unique object using only the deepest exposures.
\item {\tt FIELD/HEALPix\_allobj\_bright.fits.gz} -- Bright stars from allobj used for cross-matching between fields.
\item {\tt FIELD/HEALPix\_allobj\_xmatch.fits.gz} -- Cross-matches between SMASH and Gaia DR2 \citep{Gaia2016,GaiaDR2}, 2MASS \citep{Skrutskie2006} and ALLWISE \citep{Cutri2013} using catalogs from CDS VizieR\footnote{\url{https://vizier.u-strasbg.fr/viz-bin/VizieR}}.
\item {\tt FIELD/HEALPix\_expmap.fits.gz} --  The ``exposure" map per band.
\end{enumerate}

\noindent
More detailed descriptions of the catalogs can be found in the PHOTRED \texttt{README} file on the ftp site\footnote{\url{ftp://astroarchive.noirlab.edu/public/hlsp/smash/dr2/catalogs/README.txt}}.

The photometric and astrometric performance is the same as mentioned in \citet{Nidever2017}.  The photometric precision (calculated using multiple independent measurements of bright stars) is  10 mmag ($u$), 7 mmag ($g$), 5 mmag ($r$), 8 mmag ($i$), and 5 mmag ($z$).  The photometric accuracy measured from overlapping fields that are independently calibrated are 13 mmag ($u$), 13 mmag ($g$), 10 mmag ($r$), 12 mmag ($i$), and 13 mmag ($z$).  The median 5$\sigma$ point source photometric depths in $ugriz$ bands are (23.9, 24.8, 24.5, 24.2, 23.5) mag.  Finally, the astrometric precision of individual measurements is $\sim$20 mas for bright stars, but this number increases for fainter stars with lower S/N.  The astrometric accuracy is $\sim$2 mas per coordinates with respect to the Gaia DR1 reference frame.  

\begin{figure*}
\begin{center}
$\begin{array}{ccc}
\includegraphics[width=0.33\hsize,angle=0]{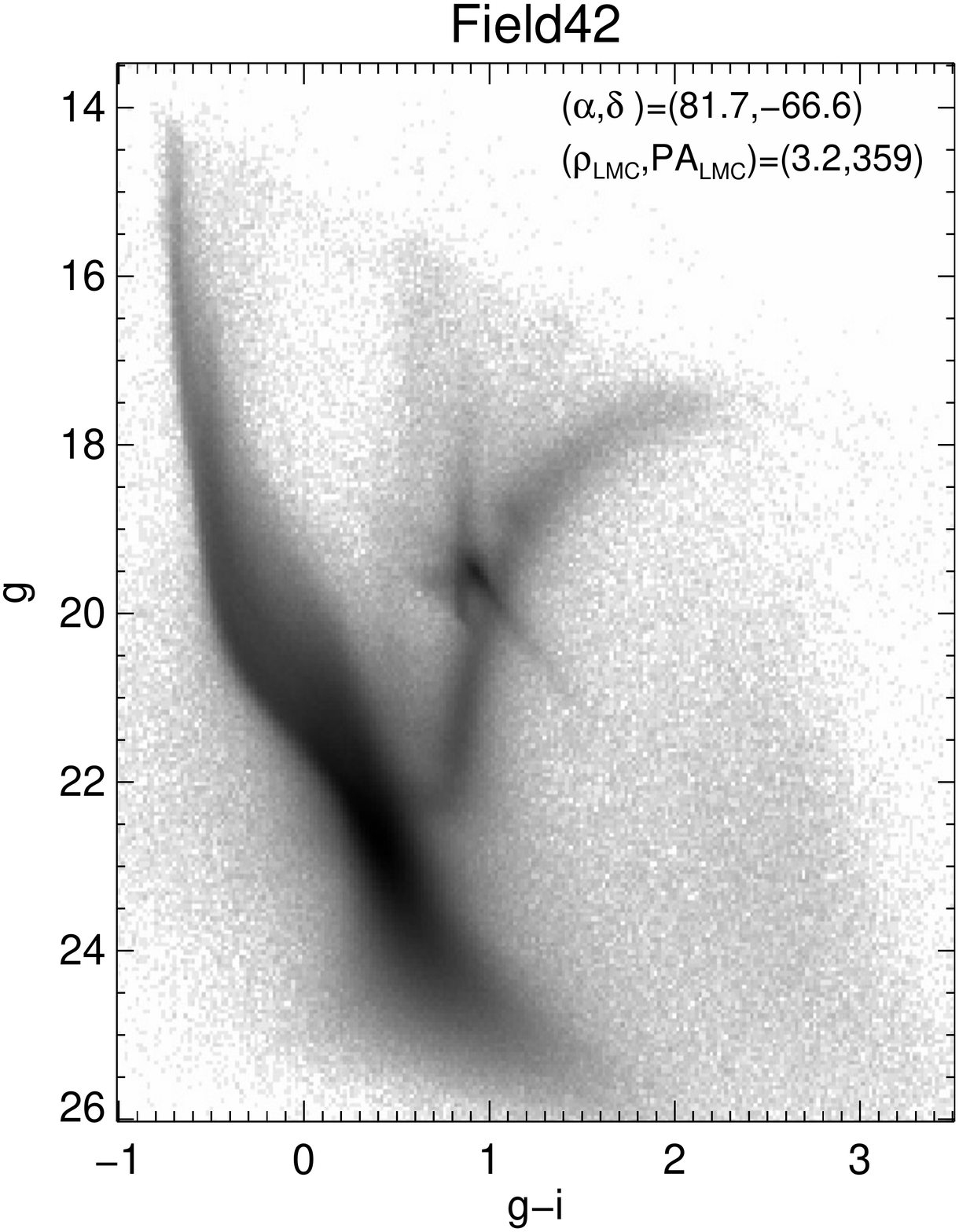}
\includegraphics[width=0.33\hsize,angle=0]{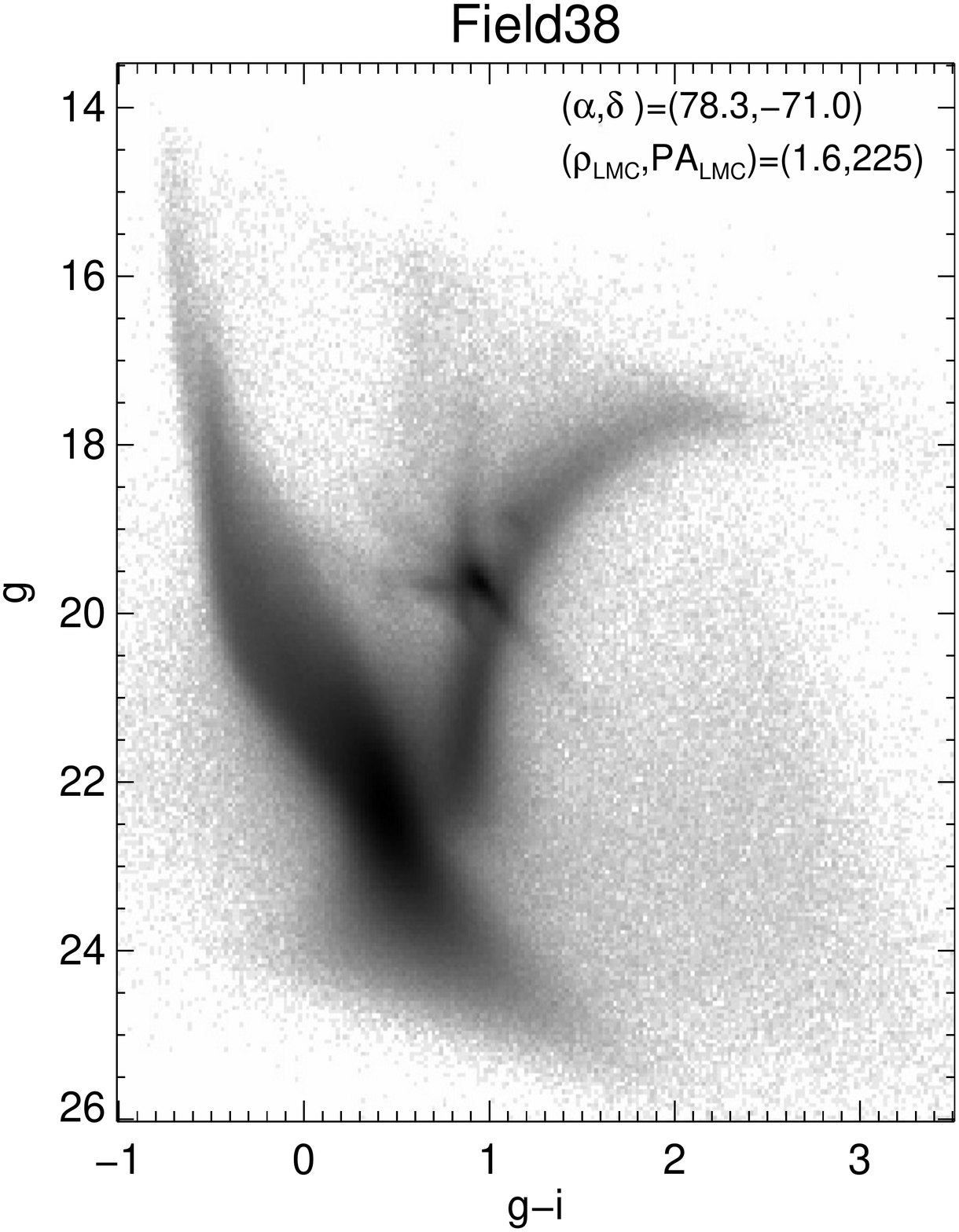}
\includegraphics[width=0.33\hsize,angle=0]{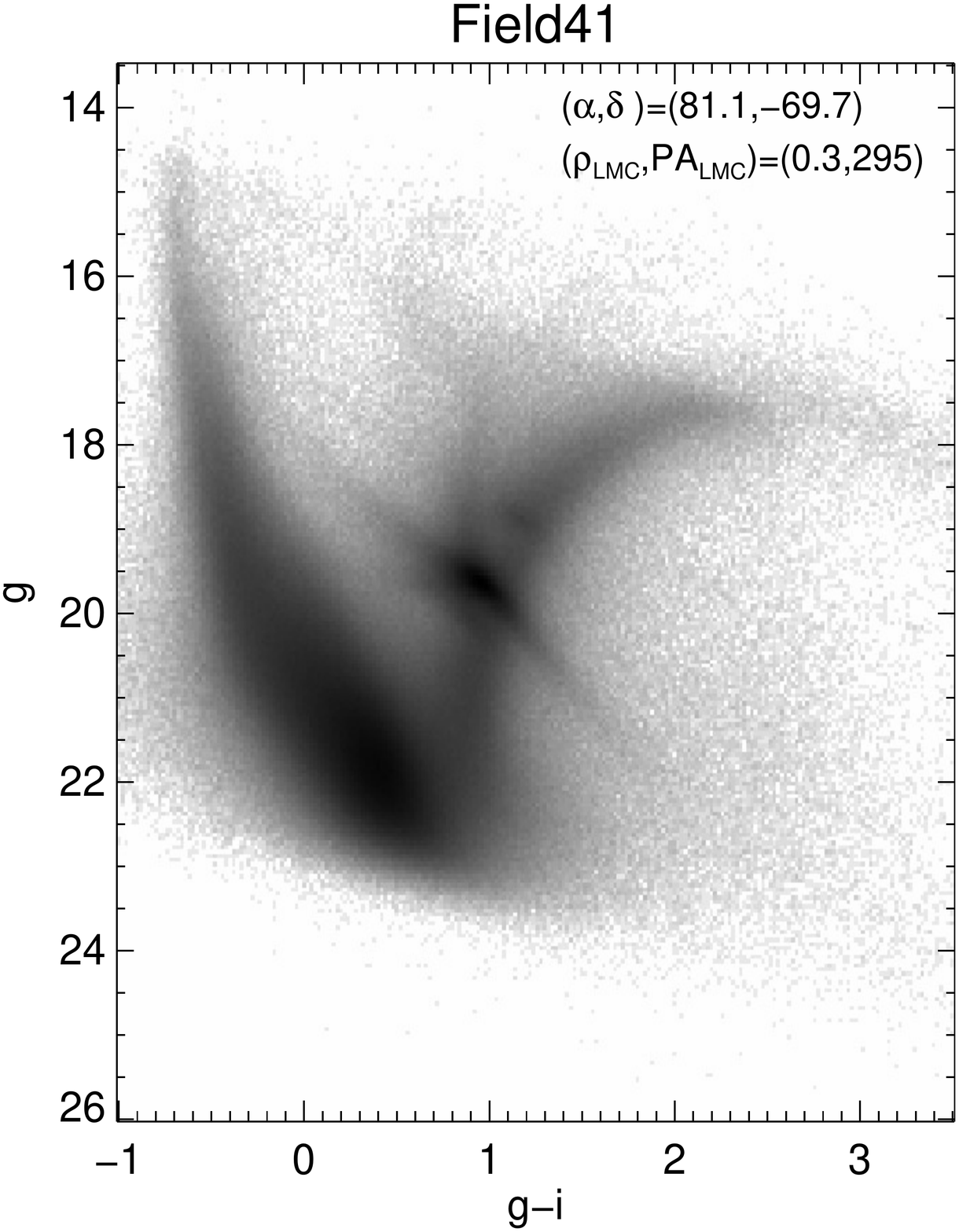} \\
\includegraphics[width=0.33\hsize,angle=0]{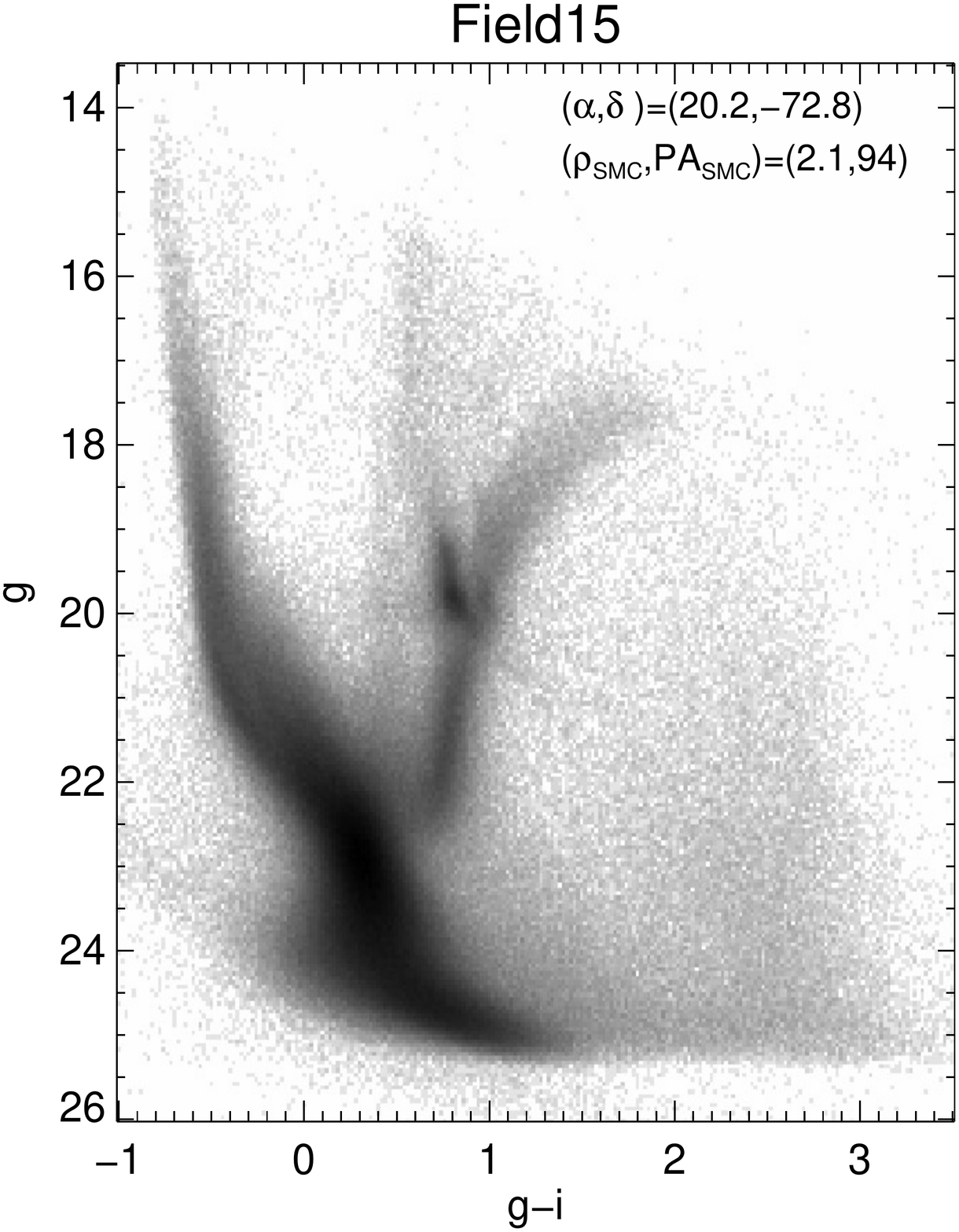}
\includegraphics[width=0.33\hsize,angle=0]{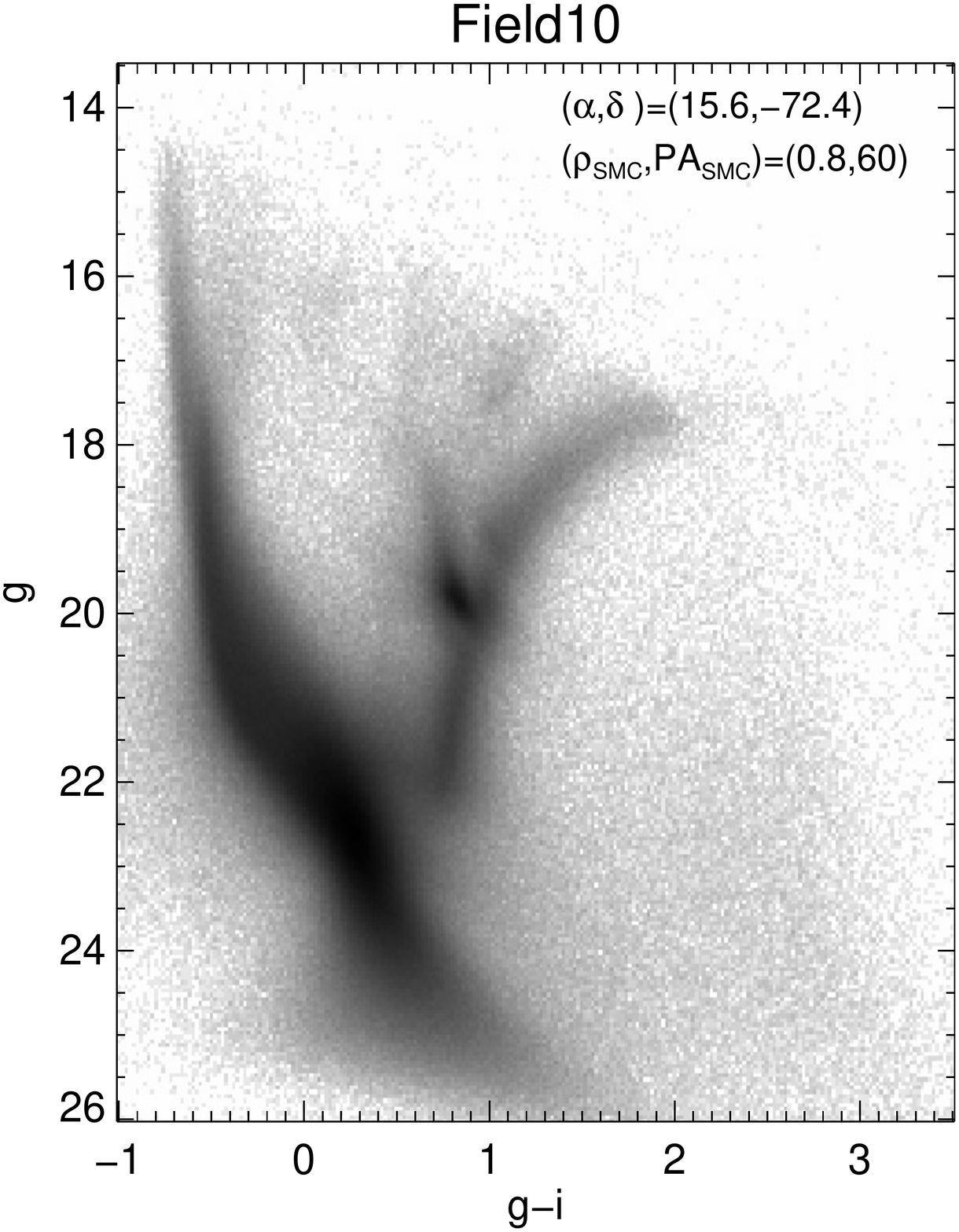}
\includegraphics[width=0.33\hsize,angle=0]{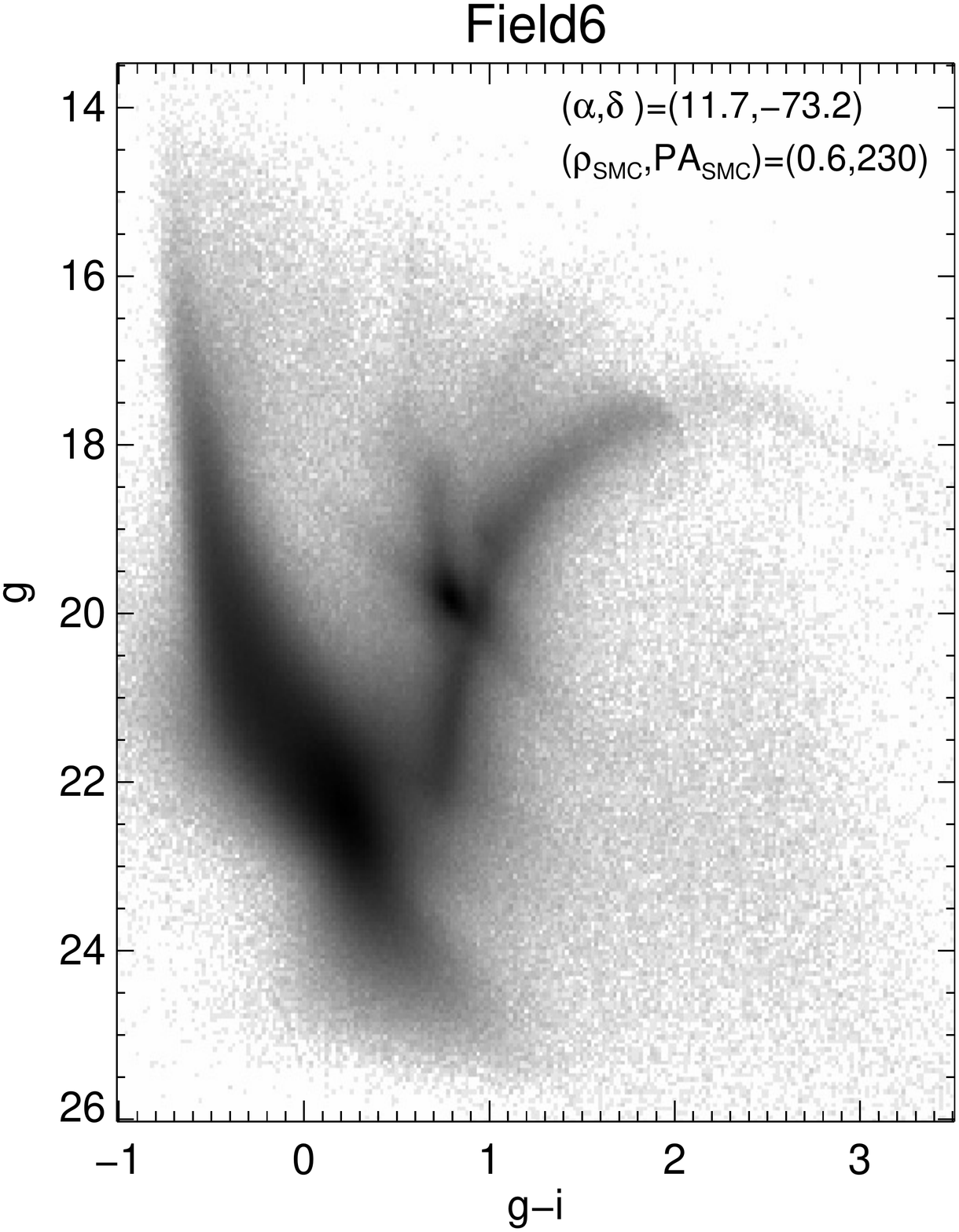}
\end{array}$
\end{center}
\caption{Example SMASH Hess diagrams of the inner LMC (top) and SMC (bottom) showing the quality of the SMASH photometry.  Field42 is in the northern LMC at a radius of 3.2\dgr and has a significant amount of reddening. Field38 is 1.6\dgr from the LMC center and has somewhat reduced depth due to crowding.  The center of the LMC is covered by Field41 which has substantial reddening and reduced depth.  Field15 is in the eastern SMC at a radius of 2.1\dgr and has very young stars from the Magellanic Bridge.  Field10 is 0.8\dgr from the SMC center and has increased reddening.  Finally, Field6 covers the center of the SMC and has reduced depth due to crowding.}
\label{fig_cmds}
\end{figure*}

Although we used the best seeing conditions during observing runs ($\approx$1\arcsec~in $i$-band) to observe the very center of the MCs, there is still appreciable crowding in these regions that somewhat reduces the depth of the photometric catalogs.
Figure \ref{fig_depth} shows the $g$-band 10$\sigma$ depth in the central contiguous regions of the MCs and the reduced depth due to crowding is clearly visible.  These effects can also be seen in the example deep SMASH color magnitude diagrams of the inner LMC and SMC in Figure \ref{fig_cmds}.

\section{Second Public Data Release}
\label{sec:dr2}

As mentioned above, the second SMASH public data release contains $\sim$4 billion measurements of $\sim$360 million objects in 197 fully-calibrated fields covering $\sim$480 square degrees and sampling $\sim$2400 deg$^2$ of the Magellanic system (Fig.\ \ref{fig_dr2map}).  Of the 61 fields released in DR1, the data for 29 fields have been reprocessed but for the other 32 fields the catalogs are identical to those in DR1 except for some additional columns (e.g., HEALPix indices, ecliptic coordinates) that we added in DR2.
As for SMASH DR1, the main data access is through the Astro Data Lab\footnote{\url{https://datalab.noao.edu/smash/smash.php}} hosted by NSF's National Optical-Infrared Astronomy Research Laboratory (NOIRLab). Access and exploration tools include a custom Data Discovery tool, database access to the catalog (via direct query or TAP service), an image cutout service, and a Jupyter notebook server with example notebooks for exploratory analysis.  The data release page also gives extensive documentation on the SMASH survey, including the observing strategy, data reduction and calibration, as well as information on the individual data products.

Images, intermediate data products, and final catalogs (in FITS binary formats) are also available through the NOIRLab High Level Data Products FTP
site\footnote{\url{ftp://astroarchive.noirlab.edu/public/hlsp/smash/dr2/}}. The raw images as well as the CP-reduced InstCal, Resampled and single-band Stacked images are
available in {\tt raw/}, {\tt instcal/}, {\tt resampled/}, and {\tt stacked/} directories, respectively (and grouped in nightly subdirectories).  Each subdirectory has a {\tt README} file that gives information about each FITS image file (e.g., exposure number, time stamp, filter, exposure time, field).  The PHOTRED-ready FITS files and other associated files (PSF, photometry catalogs, logs, etc.) as well as the multi-band stacks are available in the {\tt photred/} directory.  The final binary FITS catalogs (as described in Section \ref{subsec:catalogdescription}) are in the {\tt catalogs/} directory.  Finally, there are seven tables in the database that were populated using the FITS catalogs(with some additional columns (e.g, HEALPix indices, ecliptic coordinates): field, exposure, chip, source, object, deep, and xmatch.  The exposure map files are not loaded into the database.  A detailed description of the database schema (tables and columns) is available on the Astro Data Lab website in the Query Interface\footnote{\url{https://datalab.noao.edu/query.php}}.

\section{Science Examples}
\label{sec:examples}

SMASH is a deep photometric survey of the MCs with the goal of understanding the low density stellar structures in the periphery as well as the complex evolution and star formation history of the inner populations.  SMASH DR2 data are the deepest and most extensive covering the central regions of the Clouds in a contiguous way (see Fig.\ \ref{fig_lmc3color} and \ref{fig_smc3color}) to date and will be extremely valuable for understanding the structure and evolution of the hearts of these nearby galaxies.

\begin{figure}
\begin{center}
\includegraphics[width=0.45\textwidth]{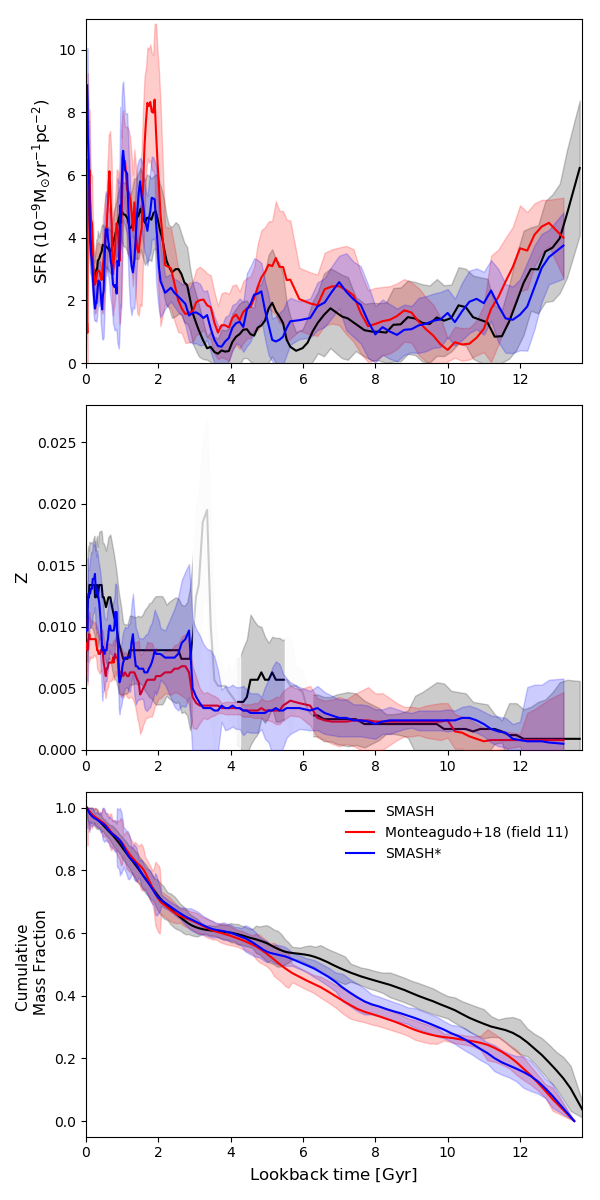}
\end{center}
\caption{Comparison between the star formation histories computed using SMASH and VIMOS \citep{Monteagudo2018} data.  We show the star formation rate as a function of time (top), the chemical enrichment (middle), and the cumulative mass fraction (bottom). We compare three different solutions: SMASH (black) using SMASH data and the set of parameters that better suits the data; Monteagudo+18 (field 11, red); and SMASH* (blue) using SMASH data and mimicking the input parameters used in \citet[][]{Monteagudo2018}. The middle panel shows with transparency the age range where a small fraction of stars has been found and thus, where a fully-reliable metallicity determination is hindered. Shaded regions represent uncertainties in the SFH recovery computed as described in \citet[][]{hidalgo2011}.}
\label{fig:sfh_example}
\end{figure}

\subsection{Star Formation Histories}
\label{subsec:sfh}

The quality and spatial coverage of the SMASH DR2 data allow us to recover deep CMDs reaching the oMSTO for the main bodies of the MCs (see Fig.~\ref{fig_cmds} as well as Figures \ref{fig_dr2map} and 11 and 12 from \citealt{Nidever2017}). This enables, among many other aspects, the derivation of spatially resolved star formation histories in the central LMC/SMC fields \citep[][Massana et al., in prep.]{RuizLara2020b}. As an example of the reliability of SMASH data to recover SFHs, and partly as a sanity check, Figure~\ref{fig:sfh_example} compares published SFHs with that obtained using SMASH data from the same region. For this comparison we have chosen field 11 from \citet{Monteagudo2018}, located on the LMC's northwest arm \citep{elyoussoufi2019} within SMASH field 37. \citet{Monteagudo2018} used the VIsible Multi-Object Spectrograph (VIMOS) on the Very Large Telescope (VLT) to obtain deep $B$ and $R$-band images in several fields within the central part of the LMC. The observations were carried out under exceptional seeing conditions (seeing $\sim$ 0.6-0.8\arcsec) and exposure times of 1832s for $B$-band and 2232s for $R$-band, resulting in completeness levels of $\sim$40$\%$ in the oMSTO region (compared to the nearly 80\% completeness we have in SMASH field 37). The quality of the VIMOS data, the similar methodology with respect to our SMASH approach, and the compatibility of their results with SFHs from deeper HST data (see figure 3 in \citealt{Monteagudo2018}) make this comparison a perfect exercise to test the reliability of SFHs recovered from SMASH data.


\begin{figure*}
\begin{center}
\includegraphics[width=1.0\textwidth]{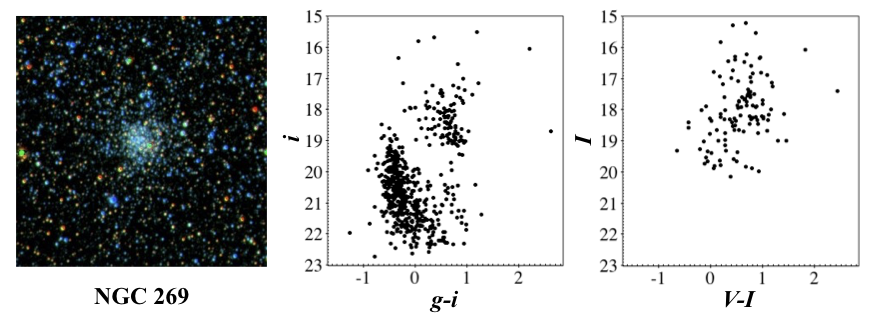}
\end{center}
\caption{Cluster NGC 269 in the SMC. Left: SMASH \textit{gri} color image, Middle: SMASH CMD, Right: MCPS CMD.}
\label{fig:clustercmd}
\end{figure*}

The methodology applied to recover SFHs from SMASH data is the result of the experience acquired over the years of members of the SMASH collaboration \citep[e.g.,][]{gallart1999, harris_zaritsky2001, aparicio2004, monelli2010, hidalgo2011, monachesi2012, meschin2014, Noel2015,bernard2018MNRAS} and it is based on the well-tested and established technique of CMD-fitting \citep[e.g.][]{gallart2005, tolstoy2009,  cignonitosi2010}. The basis of that method is extensively explained in \citet[][]{monelli2010} and \citealt[][]{bernard2018MNRAS} (see also \citealt[][]{RuizLara2020b} and Ruiz-Lara et al. in prep. for the particular case of SMASH). In short, we computed a synthetic CMD containing around 150 million stars using the BaSTI stellar evolution library (\citealt{pietrinferni2004}: solar-scaled; Reimers mass loss parameter, $\eta$ = 0.2; Kroupa initial mass function, \citealt{Kroupa2001}; binary fraction, $\beta$ = 50\%; minimum mass ratio for binaries, q = 0.1). This synthetic CMD is divided, following a grid in age and metallicity, in different single stellar populations (SSPs) that will be compared to the observed CMDs. Such comparison is carried out by counting the number of stars in small boxes in specific regions (``bundles'') within the CMD using the code {\tt THESTORM} \citep[][]{bernard2018MNRAS}. Prior to the comparison itself, the observational errors affecting SMASH data are simulated in the synthetic CMD. For that, we have developed a pipeline ({\tt FAKERED}) to compute artificial star tests for the SMASH data. The routines of this pipeline were first outlined in \citet{Choi:2018a}. {\tt FAKERED} first injects artificial stars of a large range of colors into the images, then reruns {\tt PHOTRED}, and finally tabulates which artificial stars were recovered, together with their measured magnitudes. Details about our artificial tests will be presented in Ruiz-Lara et al. (in prep.) along with the full SFHs in the deep fields observed in the LMC. The catalog of artificial stars and the completeness maps will be released in the final SMASH data release and a more detailed description of {\tt FAKERED} will be given in the accompanying paper. 

For comparison between the observed CMDs and the synthetic CMDs, we transformed the observed CMDs into the absolute plane on a star-by-star basis by taking into account spatially-varying differential reddening and distance to each field (due to the LMC's inclined disk) using the RC-based results by \citet[][]{Choi:2018a} for the LMC and Choi et al. (in prep) for the SMC, respectively. Further details about deriving the SFHs using the SMASH data are given in \citet[][]{RuizLara2020b} and will be presented in future SMASH papers. All the above-described ingredients (computation of synthetic CMDs, ``bundle'' strategy, and definition of SSPs) have been carefully chosen after extensive tests to properly deal with the SMASH data. 

The outcome of this methodology applied to the above described test region is shown in Figure~\ref{fig:sfh_example} from the analysis of SMASH (black) and VIMOS \citep[red,][]{Monteagudo2018} data. Although both datasets have been analysed following the same methodology, there are slight differences regarding the ``bundle'' definition, age and metallicity bins, and synthetic CMD computation that might affect the comparison. In order to isolate the effect of the different sets of data (SMASH vs. VIMOS), and thus, assessing the consistency of the SFHs derived with SMASH data with previous published determinations, we decided to run a different test (SMASH*, in blue) mimicking all the input parameters used in \citet[][]{Monteagudo2018}. In all cases the results are in agreement within uncertainties both in the star formation rate as a function of time (SFR(t); top panel) as well as in the chemical enrichment (middle panel). The main discrepancies are found at intermediate and old ages, although, even there, wiggles in the recovered SFR(t) showing maxima at $\sim$5, and 7 Gyr, as well as the increased SFR at ages older than $\sim$11 Gyr are quite coincident among the different solutions. As expected, the small differences found between SMASH and \citet[][]{Monteagudo2018} are somewhat minimized when comparing with this second attempt, SMASH* (especially recognisable in the cumulative mass fraction panel, bottom panel of Figure~\ref{fig:sfh_example}). 

At this point, we should emphasize that SMASH data not only provide SFH results that are compatible with data collected from other ground-based facilities. The VIMOS SFHs derived in \citet{Monteagudo2018} have also been contrasted, successfully, to SFHs derived using deeper HST data (see their Figure 3). We find, in agreement with previous studies of the SFH of the LMC disc \citep[e.g.][]{harris_zaritsky2009, meschin2014}, three clearly distinct epochs: i) an initial phase forming metal-poor, old stars (older than 11 Gyr); ii) an epoch with lower but non-negligible star formation (3.5-11 Gyr); and iii) a re-ignition of the star formation in the last 3.5 Gyr. 

In the near future, we plan to fully exploit this splendid SMASH dataset by deriving spatially resolved census of the inner region of both Clouds to tackle specific science questions for which the knowledge of SFHs is essential. The first of such works has already been published \citep[][]{RuizLara2020b}, showing compelling evidence suggesting the stability and longevity of the LMC spiral arm, in place for at least $\sim$2~Gyr.

\subsection{Star Cluster Search}
\label{subsec:clustersearch}

The star clusters of the MCs are a valuable resource, acting as testbeds and calibrators for stellar evolution models \citep[e.g.,][]{Marigo2008}, as well as providing long-lived tracers of chemical evolution and star formation \citep[e.g.,][]{Carrera2008, Piatti2013}. In the LMC, the cluster age distribution is well-known to feature a large number of clusters with ages less than $\sim$3 Gyr as well as a number of ancient globular clusters, but almost no clusters with ages between these extremes \citep{Mateo1986, Olszewski1996}.  Conversely, the field shows star formation continuing throughout this epoch, albeit at a suppressed rate \citep{Olsen1999, Holtzman1999}. 
The uniform, multi-band SMASH observations of the main bodies of the LMC and SMC provide the necessary depth and image quality (1.1\arcsec~average seeing) to enhance star cluster searches and catalogs, and to provide a basis for comparing cluster and field star formation histories from the same data. 

Previous star cluster searches and catalogs are the result of either heterogeneous data that is the combination of many individual studies spanning decades \citep{Bica2008, Bica2019} or poorer data from MCPS \citep{hill_zaritsky2006, werchan_zaritsky2011} or OGLE \citep{Pietrzynski1999}. The SMASH dataset provides the opportunity to re-derive a star cluster catalog for the MCs using uniform data and cluster detection techniques which should be more spatially complete than previous searches. This new effort aims to overcome shortcomings of recent algorithmic detection efforts \citep[see][]{Bitsakis2017, Bitsakis2018, Piatti18_LMC, Piatti18_SMC} by conducting a well-characterized crowdsourced visual search of the SMASH imaging.  This technique continues the long tradition of visual cluster identification across the Local Group, and builds on the legacy of crowdsourced cluster identification in the Andromeda Galaxy \citep{Johnson2015}, which demonstrated the capability and robustness of this methodology.

The crowdsourced search of the SMASH data is currently underway via the Local Group Cluster Search\footnote{\url{clustersearch.org}}, a citizen science project hosted on the Zooniverse\footnote{\url{zooniverse.org}} platform.  In addition to producing a uniformly derived catalog, this effort will also produce the first robust star cluster catalog completeness determination for the MC cluster population through its use of synthetic cluster tests included in the imaging.  The ability to model observational completeness for the cluster sample will significantly improve population level analyses.

Beyond the updated cluster catalog, the SMASH photometric data provide deep cluster CMDs
that suffer from less crowding than previous galaxy-wide photometric catalogs, enabling significant improvements to cluster CMD fitting \citep[e.g.,][]{Glatt2010} that extend to old cluster ages.  In Figure~\ref{fig:clustercmd}, we demonstrate the improvement of SMASH cluster CMDs over previous MCPS-based data.  Together, the catalog and CMD improvements from SMASH will usher in a new wave of population-wide analyses of MC clusters.

\subsection{Photometric Metallicities}
\label{subsec:photmetals}

\begin{figure}[t]
\includegraphics[width=1.0\hsize,angle=0]{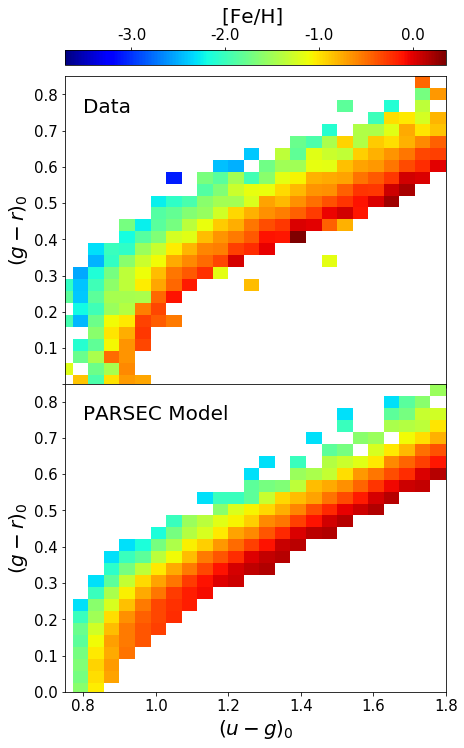}
\caption{The SMASH MSTO range in color-color space. (Top) Binned SDSS and LAMOST spectroscopic median metallicity as a function of $(g-r)_0$ vs.\ $(u-g)_0$ in SMASH colors, with bin sizes of 0.033 dex in $(g-r)_0$ and 0.043 dex in $(u-g)_0$. (Bottom) Binned PARSEC isochrone metallicity with the same binning.}
\label{fig_photmetals}
\end{figure}

The SMASH $u$-band can be used to obtain photometric metallicities by adopting the methods used by \citet{Ivezic2008} for SDSS data and \citet{ibata2017} for the Canada-France Imaging Survey $u$-band together with SDSS and Pan-STARRS $g$, $i$, and $r$ photometry.
Here we introduce our analysis of SMASH dwarf stars and focus on the Magellanic MSTO population, but we also plan to investigate measuring metallicities for Gaia DR2 proper motion-selected Magellanic RGB stars. 
We recalibrate their methods with SMASH $u$, $g$, and $r$ photometry and two spectroscopic surveys: SDSS \citep{Ahn2014} and the Large sky Area Multi-Object Spectroscopic Telescope LAMOST \citep[LAMOST;][]{Luo2015}. There are $\sim$4,000 stars in SMASH standard star fields that have SDSS or LAMOST spectroscopic metallicities. 
We select $\sim$3,000 dwarfs using a spectroscopic classification of 3 $<$ $\log{g}$ $<$ 5. We find sorting our calibration stars by $u-g$ color versus $g-r$ color is optimal for tracing metallicity of SMASH dwarf stars; this trend is supported by the literature for similar MW dwarf stars \citep{Ivezic2008,ibata2017}. The binned calibration sample of the dwarfs with both spectroscopic metallicities and SMASH photometry is shown in the top panel of Figure \ref{fig_photmetals}. We also construct a set of PARSEC isochrones \citep{Bressan2012,marigo2017} of age 3 Gyrs with a minimum and maximum metallicity value matching those of our calibration sample; the binned isochrone tracks are shown in the bottom panel of Figure \ref{fig_photmetals}. We use two different techniques to calculate a star's photometric metallicity with its SMASH $(u-g)_0$ and $(g-r)_0$ colors: (1) the mean of the five nearest metallicity points in the spectroscopic mean metallicity map (top panel of Figure \ref{fig_photmetals}), and (2) the mean of the five nearest metallicity points in the PARSEC isochrone mean metallicity map (bottom panel of Figure \ref{fig_photmetals}).  These two methods produce very similar metallicity results.

\begin{figure}[t]
\includegraphics[width=0.47\textwidth]{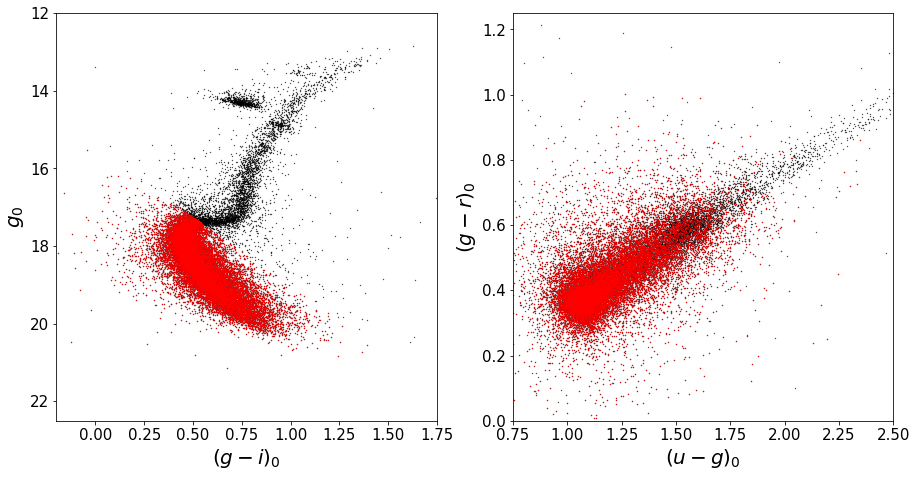}
\caption{The MW cluster 47 Tucanae in the SMASH data. (Left) CMD of the cluster member stars with the selected main sequence population in red. (Right) Color--color diagram of cluster member stars.}
\label{fig_47tuc}
\end{figure}

We test our photometric metallicity technique on two prominent Milky Way clusters in our SMASH data: 47 Tucanae and NGC 362.
47 Tuc is contained within SMASH field 4 while NGC 362 is contained within SMASH field 9. Figure \ref{fig_47tuc} shows the stellar density and CMD (with selected main-sequence stars in red) of 47 Tuc.  We run the main-sequence populations of both clusters through our method.  We obtain a metallicity of [Fe/H]=$-$0.76 for 47 Tuc which is consistent with the literature value of [Fe/H]=$-$0.76 \citep{Harris1997}.  For NGC 362, we obtain [Fe/H]=$-$1.03 which is slightly discrepant from the literature value of [Fe/H]=$-$1.16 \citep{Harris1997}.  As an additional check, we have used the high-resolution spectrosopic APOGEE-2 data from SDSS DR16 \citep{Ahumada2020} --- which targeted both of these clusters --- to measure median metallicities of [Fe/H]=$-$0.73 for 47 Tuc and [Fe/H]=$-$1.09 for NGC 362.  These results indicate that our SMASH photometric metallicities are accurate at the 0.02--0.06 dex level at metallicities of [Fe/H]$\approx$$-$1. 

Although these results demonstrate the veracity of our general technique, we acknowledge that it is more challenging to apply our method to the Magellanic MSTO populations.  These stars are several magnitudes fainter than the cluster main-sequence stars and have larger photometric uncertainties.  In addition, reddening can be an important issue for the $u$-band which is vital to our study. To combat these issues, we only study SMASH fields on the outer main--bodies of the MCs and the Magellanic periphery where dust extinction is dominated by the MW and \citet{Schlegel1998} dust corrections can be reliably applied. Also, we will take advantage of the thousands of MSTO stars per field and use statistically robust methods to represent the average photometric metallicity.
The results will be presented in an upcoming paper (Miller et al., in prep.) which will also investigate the effects of age and binarity on the photometric metallicities.


%

SMASH photometric metallicities that we measure in the Magellanic periphery will ultimately help shed light on the origin of the outer, low surface brightness structure of the Magellanic Clouds \citep[e.g.,][]{Belokurov2019,Nidever2019a}.  A ``classical'' halo of accreted satellites should be quite metal-poor ([Fe/H] $\approx$ $-$2) as the satellites of the MCs would be quite low-mass and metal-poor.  An ``in situ'' or puffed-up disk origin would be relatively more metal-rich and similar to the outer disk of the LMC and spheroidal component of the SMC ([Fe/H] $\approx$ $-$1).  A more in-depth discussion of our photometric metallicity technique as well as the metallicity results in the outskirts of the Magellanic Clouds will be presented in Miller et al.\  (in prep.).

\section{Summary \& Future Outlook}

Except for the artificial star catalogs, SMASH DR2 marks the completion of the survey's public data releases, but multiple science investigations introduced in Sec. \ref{sec:examples} and beyond will continue.  The SMASH survey has the potential to revolutionize our understanding of the stellar populations inside and to the very outskirts of two canonical examples of dwarf galaxies, the SMC and LMC.

DECam surveys continue to explore the Magellanic Clouds and their surroundings, extending the investigations begun by SMASH.  Of particular note, the legacy of the SMASH survey is continued as part of the DECam Local Volume Exploration (DELVE)\footnote{\url{https://delve-survey.github.io/}} survey.  This three-year, three-component NOIRLab observing program includes a Magellanic Clouds periphery survey (DELVE-MC) that uses DECam to obtain contiguous $gri$ coverage across 1000 deg$^2$ to a $g$-band depth of $\sim$24.2 mag.

\acknowledgments

Y.C., E.F.B., and A.M. acknowledge support from NSF grant AST 1655677.
A.D., C.G., T.R.L. and M.M. acknowledge support by the Spanish Ministry of Economy and Competitiveness (MINECO) under the grants AYA2014-56795-P and AYA2017-89076-P as well as AYA2016-77237-C3-1-P. T.R.L. has support from a Spinoza grant (NWO) awarded to A. Helmi and acknowledges support by a MCIU Juan de la Cierva - Formaci\'on grant (FJCI-2016-30342). C.P.M.B. and M.-R.L.C. acknowledge support from the European Research Council (ERC) under the European Union’s Horizon 2020 research and innovation programme (grant agreement No 682115). A.M. acknowledges support from FONDECYT Regular 1181797.
R.R.M. acknowledges partial support from project BASAL AFB-$170002$ as well as FONDECYT project N$^{\circ}1170364$.
D.M.D. acknowledges financial support from the State Agency for Research of the Spanish MCIU through the
``Center of Excellence Severo Ochoa'' award to the Instituto de Astrofísica de Andalucía (SEV-2017-0709).
Image processing: Travis Rector (University of Alaska Anchorage), Mahdi Zamani \& Davide de Martin.  We thank the anonymous referee for useful comments that improved the manuscript.

Based on observations at Cerro Tololo Inter-American Observatory, NSF's National Optical-Infrared Astronomy Research Laboratory (NOAO Prop. ID: 2013A-0411 and 2013B-0440; PI: Nidever), which is operated by the Association of Universities for Research in Astronomy (AURA) under a cooperative agreement with the National Science Foundation.
IRAF is distributed by the National Optical Astronomy Observatory, which is operated by the Association of Universities for Research in Astronomy (AURA) under a cooperative agreement with the National Science Foundation.
This project used data obtained with the Dark Energy Camera (DECam), which was constructed by the Dark Energy Survey (DES) collaboration. Funding for the DES Projects has been provided by the U.S. Department of Energy, the U.S. National Science Foundation, the Ministry of Science and Education of Spain, the Science and Technology Facilities Council of the United Kingdom, the Higher Education Funding Council for England, the National Center for Supercomputing Applications at the University of Illinois at Urbana-Champaign, the Kavli Institute of Cosmological Physics at the University of Chicago, Center for Cosmology and Astro-Particle Physics at the Ohio State University, the Mitchell Institute for Fundamental Physics and Astronomy at Texas A\&M University, Financiadora de Estudos e Projetos, Funda\c{c}\~ao Carlos Chagas Filho de Amparo, Financiadora de Estudos e Projetos, Funda\c{c}\~ao Carlos Chagas Filho de Amparo \`a Pesquisa do Estado do Rio de Janeiro, Conselho Nacional de Desenvolvimento Cient\'ifico e Tecnol\'ogico and the Minist\'erio da Ci\^encia, Tecnologia e Inova\c{c}\~ao, the Deutsche Forschungsgemeinschaft and the Collaborating Institutions in the Dark Energy Survey. The Collaborating Institutions are Argonne National Laboratory, the University of California at Santa Cruz, the University of Cambridge, Centro de Investigaciones En\'ergeticas, Medioambientales y Tecnol\'ogicas-Madrid, the University of Chicago, University College London, the DES-Brazil Consortium, the University of Edinburgh, the Eidgen\"ossische Technische Hochschule (ETH) Z\"urich, Fermi National Accelerator Laboratory, the University of Illinois at Urbana-Champaign, the Institut de Ci\`encies de l'Espai (IEEC/CSIC), the Institut de F\'isica d'Altes Energies, Lawrence Berkeley National Laboratory, the Ludwig-Maximilians Universit\"at M\"unchen and the associated Excellence Cluster Universe, the University of Michigan, the National Optical Astronomy Observatory, the University of Nottingham, the Ohio State University, the University of Pennsylvania, the University of Portsmouth, SLAC National Accelerator Laboratory, Stanford University, the University of Sussex, and Texas A\&M University.

This work has made use of data from the European Space Agency (ESA) mission {\it Gaia} (\url{https://www.cosmos.esa.int/gaia}), processed by the {\it Gaia} Data Processing and Analysis Consortium (DPAC,
\url{https://www.cosmos.esa.int/web/gaia/dpac/consortium}). Funding for the DPAC has been provided by national institutions, in particular the institutions participating in the {\it Gaia} Multilateral Agreement.

This publication makes use of data products from the Two Micron All Sky Survey, which is a joint project of the University of Massachusetts and the Infrared Processing and Analysis Center/California Institute of Technology, funded by the National Aeronautics and Space Administration and the National Science Foundation.

\bibliography{ref_og.bib}

\begin{deluxetable}{lccccccc}
\centering 
\tablecaption{SMASH Fields Table}
\tablecolumns{8}
\tablewidth{500pt}
\tablehead{
  \colhead{Number} & \colhead{Name} & \colhead{RAJ2000} & \colhead{DEJ2000} & \colhead{RADEG} &
  \colhead{DEDEG} & \colhead{L$_{\rm MS}$\tablenotemark{1}} & \colhead{B$_{\rm MS}$\tablenotemark{1}} \\
  \colhead{} & \colhead{} & \colhead{(hms)} &  \colhead{(hms)} &  \colhead{(deg)} &  \colhead{(deg)}
  &  \colhead{(deg)} &  \colhead{(deg)} \\
}

\startdata
   1 &  0010-6947 &   00:10:19.87 &  $-$69:47:40.56 &     2.58282 &  $-$69.794600 &   $-$19.56937 &   $-$13.84173 \\
   2 &  0018-7705 &   00:18:57.90 &  $-$77:05:00.23 &     4.74128 &  $-$77.083400 &   $-$12.11547 &   $-$15.01799 \\
   3 &  0023-7358 &   00:23:19.08 &  $-$73:58:08.04 &     5.82950 &  $-$73.968900 &   $-$15.14808 &   $-$13.94466 \\
   4 &  0024-7223 &   00:24:56.64 &  $-$72:23:08.15 &     6.23604 &  $-$72.385600 &   $-$16.67737 &   $-$13.38564 \\
   5 &  0044-7137 &   00:44:06.76 &  $-$71:37:45.84 &    11.02820 &  $-$71.629400 &   $-$16.91433 &   $-$11.74021 \\
   6 &  0044-7313 &   00:44:32.54 &  $-$73:13:31.79 &    11.13560 &  $-$73.225500 &   $-$15.38135 &   $-$12.28828 \\
   7 &  0045-7448 &   00:45:03.28 &  $-$74:48:14.76 &    11.26370 &  $-$74.804100 &   $-$13.85882 &   $-$12.82186 \\
   8 &  0050-8228 &   00:50:34.51 &  $-$82:28:26.40 &    12.64380 &  $-$82.474000 &    $-$6.36406 &   $-$15.28088 \\
   9 &  0101-7043 &   01:01:27.40 &  $-$70:43:05.51 &    15.36420 &  $-$70.718200 &   $-$17.19874 &   $-$10.09455 \\
  10 &  0103-7218 &   01:03:36.31 &  $-$72:18:54.36 &    15.90130 &  $-$72.315100 &   $-$15.66017 &   $-$10.63165
\enddata
\tablenotetext{0}{Full table is available in the electronic version of this paper.}
\tablenotetext{1}{Magellanic Stream coordinates defined in \citet{Nidever2008}.}
\label{table_smashfields}
\end{deluxetable}

\begin{center}
\begin{deluxetable}{llll}
\tablecaption{SMASH DECam and 0.9-m Observing Runs}
\tablecolumns{4}
\tablewidth{590pt}
\tablehead{
\colhead{Date (nights)} & \colhead{Telescope} & \colhead{Source} & \colhead{Comments}
}
\startdata
\sidehead{Pre-Survey}
\hline4-m
Dec 11+12, 2012 (2)     & 4-m    & Shared Risk  &  5 pilot fields \\
Mar 17--20, 2013 (4)      & 4-m  & 2013A-0411  &  23 fields ($griz$)  \\
Aug 8+9, 2013 (2 part)      &  4-m    & Time from Saha Bulge project & clear, 3 fields\hspace{4.5cm}  \\
\hline
\sidehead{Survey Year 1}
\hline
Sep 7--10, 2013 (4)  &  0.9-m  & NOAO survey & bad weather, no data  \\
Sep 11--13, 2013 (3)  &   0.9-m  & Bought from SMARTS & bad weather, no data \\
Oct 21+22, 2013 (2 part)  &  0.9-m  & Makeup for Sep 11--13 &  bad weather, no data \\ 
Jan 5--7, 2014 (3)     &  4-m  & NOAO survey  & 0.5 night lost, 10 fields \\
Jan 12+13, 2014 (2)    &  0.9-m  & Makeup for Oct 21+22  & 2 nights photometric, 4 fields calibrated \\
Jan 19+20, 2014 (2 half)    &  4-m    & DD time  & clear, $riz$ for 6 fields\\
Jan 21--28, 2014 (8 half) &  4-m    & Chilean time  &  1 half night lost, 4 fields, 9 partials \\
Jan 29+30, 2014 (2 half)    &  4-m    & DD time  &  clear, $ug$ for 8 pre-survey fields\\ 
Feb 13, 2014  (1 part)  &  4-m    & Engineering  & clear, $riz$ for 6 fields \\
Feb 14--23, 2014 (10)  &  0.9-m  & NOAO survey &  9 nights photometric, 30 fields calibrated\\
May 27--June 2, 2014 (7)  & 4-m & NOAO survey & lost 1 night, 21 fields observed, $ug$ for \\
& & & 13 pre-survey fields, 3 extra fields \\
\hline
\sidehead{Survey Year 2}
\hline
Sep 25--Oct 1, 2014 (7)  & 0.9-m  & NOAO survey &  1 night photometric, 11 fields calibrated \\
Oct 11--12, 2014 (2)  &  4-m  &  Engineering  &  some globular cluster calibration data  \\ 
Nov 21--23, 2014 (3)  &  4-m  &   NOAO survey &  12 LMC/SMC main-body fields \\
Dec 17--18, 2014 (2)  &  4-m  &   NOAO survey &  8 LMC/SMC main-body fields \\
Mar 13--18, 2015 (5) &  4-m  &   NOAO survey &  mostly clear, 21 finished, 4 partials \\
Mar 30--31, 2015 (2)   &  4-m    & DD time  & deep \& high-cadence data of Hydra II \\ 
Apr 26--Mar 2, 2015 (7) & 0.9-m &  NOAO survey & 4.5 nights photometric, 48 fields calibrated \\
\hline
\sidehead{Survey Year 3}
\hline
Oct 25+27, 2015 (2)    &  4-m       & DD time &  bad weather, no data \\ 
Nov 9, 2015 (1)   &  4-m &  NOAO survey  &  clear, 4 fields  \\
Nov 23, 2015 (1)        &  4-m       & DD time &  bad weather, long $riz$ for 2 fields \\
Nov 27--29, 2015 (3)    &  0.9-m     &  Chilean time & 9 fields calibrated \\
Dec 5+6, 2015 (2)  & 4-m &  NOAO survey & 8 fields, 7 are LMC/SMC main-body \\
Jan 1--6, 2016 (6)  &  4-m  &   NOAO survey   &  4 nights lost, 3 finished, 2 partials  \\
Feb 13--18, 2016 (6) & 4-m  &  NOAO survey  &  40 shallow LMC fields, 18 long fields \\
May 8--12, 2016 (5) & 4-m &  NOAO survey & bad weather, no data \\
\hline
\sidehead{Survey Year 4 -- Extension}
\hline
Oct 29--31, 2016 (3)    &  4-m       & NOAO survey &  0.5 night lost, 8 fields \\
Mar 8, 2017 (0.5)  &  4-m   & DD time &  clear, calibration data for 25 fields \\ 
Aug 4, 2017 (partial)  &  4-m   & DD time &  clear, calibration data for 3 fields \\
Nov 29, 2017 (1)  &  4-m   & Engineering &  clear, calibration data for 3 fields
\enddata
\label{table_observations}
\end{deluxetable}
\end{center}

\end{document}

%% file: shorthand.tex
\newcommand{\dgr}{$^{\circ}~$}

\newcommand{\hi}{H{\footnotesize I} }

\defcitealias{PriceWhelan2018}{Paper I}

\newcommand{\bs}[1]{\boldsymbol{#1}}
\renewcommand{\vec}[1]{\bs{#1}}


%% file: authors.tex
\correspondingauthor{David L. Nidever}
\email{dnidever@montana.edu}


\author[0000-0002-1793-3689]{David L. Nidever}
\affiliation{Department of Physics, Montana State University, P.O. Box 173840, Bozeman, MT 59717-3840}
\affiliation{NSF's National Optical-Infrared Astronomy Research Laboratory, 950 North Cherry Ave, Tucson, AZ 85719}

\author[0000-0002-7134-8296]{Knut Olsen}
\affiliation{NSF's National Optical-Infrared Astronomy Research Laboratory, 950 North Cherry Ave, Tucson, AZ 85719}

\author[0000-0003-1680-1884]{Yumi Choi}
\affiliation{Space Telescope Science Institute, 3700 San Martin Drive, Baltimore, MD 21218}
\affiliation{Department of Physics, Montana State University, P.O. Box 173840, Bozeman, MT 59717-3840}
\affiliation{Steward Observatory, University of Arizona, 933 North Cherry Avenue, Tucson, AZ 85721, USA}

\author[0000-0001-6984-4795]{Tomas Ruiz-Lara}
\affiliation{Kapteyn Astronomical Institute, University of Groningen, Landleven 12, 9747 AD Groningen, The Netherlands}
\affiliation{Instituto de Astrof\'{i}sica de Canarias, La Laguna, Tenerife, Spain}
\affiliation{Departamento de Astrof\'{i}sica, Universidad de La Laguna, Tenerife, Spain}

\author{Amy E. Miller}
\affiliation{Leibniz-Institut f\"{u}r Astrophysik Potsdam (AIP), An der Sternwarte 16, 14482 Potsdam Germany}
\affiliation{Department of Physics, Montana State University, P.O. Box 173840, Bozeman, MT 59717-3840}

\author[0000-0001-6421-0953]{L. Clifton Johnson}
\affiliation{Center for Interdisciplinary Exploration and Research in Astrophysics (CIERA) and Department of Physics and Astronomy, Northwestern University, 1800 Sherman Ave, Evanston, IL 60201}


\author[0000-0003-0642-6558]{Cameron P. M. Bell}
\affiliation{Leibniz-Institut f\"{u}r Astrophysik Potsdam (AIP), An der Sternwarte 16, 14482 Potsdam Germany}

\author[0000-0002-8622-4237]{Robert D. Blum}
\affiliation{NSF's National Optical-Infrared Astronomy Research Laboratory/LSST, 950 North Cherry Ave, Tucson, AZ 85719}

\author[0000-0002-6797-696X]{Maria-Rosa L. Cioni}
\affiliation{Leibniz-Institut f\"{u}r Astrophysik Potsdam (AIP), An der Sternwarte 16, 14482 Potsdam Germany}

\author[0000-0001-6728-806X]{Carme Gallart}
\affiliation{Instituto de Astrof\'{i}sica de Canarias, La Laguna, Tenerife, Spain}
\affiliation{Departamento de Astrof\'{i}sica, Universidad de La Laguna, Tenerife, Spain}

\author[0000-0003-2025-3147]{Steven R. Majewski}
\affiliation{Department of Astronomy, University of Virginia, Charlottesville, VA 22904, USA}

\author[0000-0002-1349-202X]{Nicolas F. Martin}
\affiliation{Universit\'e de Strasbourg, CNRS, Observatoire astronomique de Strasbourg, UMR 7550, F-67000 Strasbourg, France}
\affiliation{Max-Planck-Institut f\"ur Astronomie, K\"onigstuhl 17, D-69117 Heidelberg, Germany}

\author[0000-0002-8093-7471]{Pol Massana}
\affiliation{Department of Physics, University of Surrey, Guildford, GU2 7XH, UK}

\author[0000-0003-2325-9616]{Antonela Monachesi}
\affiliation{Instituto de Investigación Multidisciplinar en Ciencia y Tecnología, Universidad de La Serena, Raúl Bitrán 1305, La Serena, Chile}
\affiliation{Departamento de Astronom\'ia, Universidad de La Serena, Av. Juan Cisternas 1200 Norte, La Serena, Chile}

\author[0000-0002-8282-469X]{Noelia E. D. No\"el}
\affiliation{Department of Physics, University of Surrey, Guildford, GU2 7XH, UK}

\author{Joanna D. Sakowska}
\affiliation{Department of Physics, University of Surrey, Guildford, GU2 7XH, UK}

\author[0000-0001-7827-7825]{Roeland P. van der Marel}
\affiliation{Space Telescope Science Institute, 3700 San Martin Drive, Baltimore, MD 21218}
\affiliation{Center for Astrophysical Sciences, Department of Physics \& Astronomy, Johns Hopkins University, Baltimore, MD 21218, USA}
  
\author[0000-0002-7123-8943]{Alistair R. Walker}
\affiliation{Cerro Tololo Inter-American Observatory, NSF's National Optical-Infrared Astronomy Research Laboratory, Casilla 603, La Serena, Chile}

\author[0000-0002-5177-727X]{Dennis Zaritsky}
\affiliation{Steward Observatory, University of Arizona, 933 North Cherry Avenue, Tucson AZ, 85721}


\author[0000-0002-5564-9873]{Eric F. Bell}
\affiliation{Department of Astronomy, University of Michigan, 1085 S. University Ave., Ann Arbor, MI 48109-1107, USA}

\author[0000-0001-6959-4546]{Blair C. Conn}
\affiliation{Research School of Astronomy \& Astrophysics, Mount Stromlo Observatory, Cotter Road, Weston Creek, ACT 2611, Australia}
\affiliation{Presently: Analytics, IAG Limited, Level 4, Darling Park Tower 2, 201 Sussex Street, Sydney, NSW 2000,Australia}

\author[0000-0001-5486-2747]{Thomas J. L. de Boer}
\affiliation{Institute for Astronomy, University of Hawaii, 2680 Woodlawn Drive Honolulu, HI 96822-1897}

\author[0000-0002-4588-6517]{Robert A. Gruendl}
\affiliation{National Center for Supercomputing Applications, 1205 West Clark St., Urbana, IL 61801, USA}
\affiliation{Department of Astronomy, University of Illinois, 1002 West Green St., Urbana, IL 61801, USA}

\author[0000-0001-5292-6380]{Matteo Monelli}
\affiliation{Instituto de Astrof\'{i}sica de Canarias, La Laguna, Tenerife, Spain}
\affiliation{Departamento de Astrof\'{i}sica, Universidad de La Laguna, Tenerife, Spain}

\author[0000-0002-0810-5558]{Ricardo R. Mu\~noz}
\affiliation{Departamento de Astronom\'ia, Universidad de Chile, Camino del Observatorio 1515, Las Condes, Santiago, Chile}

\author[0000-0002-6839-4881]{Abhijit Saha}
\affiliation{NSF's National Optical-Infrared Astronomy Research Laboratory, 950 North Cherry Ave, Tucson, AZ 85719}

\author[0000-0003-4341-6172]{A. Katherina Vivas}
\affiliation{Cerro Tololo Inter-American Observatory, NSF's National Optical-Infrared Astronomy Research Laboratory, Casilla 603, La Serena, Chile}


\author[0000-0002-8722-225X]{Edouard Bernard}
\affiliation{Université Côte d'Azur, OCA, CNRS, 06304, Nice, Lagrange, France}

\author[0000-0003-0715-2173]{Gurtina Besla}
\affiliation{Steward Observatory, University of Arizona, 933 North Cherry Avenue, Tucson AZ, 85721}

\author{Julio A. Carballo-Bello}
\affiliation{Instituto de Alta Investigaci\'on, Universidad de Tarapac\'a, Casilla 7D, Arica, Chile}

\author{Antonio Dorta}
\affiliation{Instituto de Astrof\'{i}sica de Canarias, La Laguna, Tenerife, Spain}
\affiliation{Departamento de Astrof\'{i}sica, Universidad de La Laguna, Tenerife, Spain}

\author[0000-0003-3835-2231]{David Martinez-Delgado}
\affiliation{Instituto de Astrof\'isica de Andaluc\'ia, CSIC, E-18080, Granada, Spain}

\author{Alex Goater}
\affiliation{Department of Physics, University of Surrey, Guildford, GU2 7XH, UK}

\author[0000-0001-7633-3985]{Vadim Rusakov}
\affiliation{Cosmic Dawn Center (DAWN)}
\affiliation{Niels Bohr Institute, University of Copenhagen, Lyngbyvej 2, DK-2100 Copenhagen {\O}, Denmark}

\author[0000-0003-1479-3059]{Guy S. Stringfellow}
\affiliation{Center for Astrophysics and Space Astronomy, University of Colorado, 389 UCB, Boulder, CO, 80309-0389, USA}

%% file: main.bbl
\begin{thebibliography}{}
\expandafter\ifx\csname natexlab\endcsname\relax\def\natexlab#1{#1}\fi
\providecommand{\url}[1]{\href{#1}{#1}}
\providecommand{\dodoi}[1]{doi:~\href{http://doi.org/#1}{\nolinkurl{#1}}}
\providecommand{\doeprint}[1]{\href{http://ascl.net/#1}{\nolinkurl{http://ascl.net/#1}}}
\providecommand{\doarXiv}[1]{\href{https://arxiv.org/abs/#1}{\nolinkurl{https://arxiv.org/abs/#1}}}

\bibitem[{{Ahn} {et~al.}(2014){Ahn}, {Alexandroff}, {Allende Prieto}, {Anders},
  {Anderson}, {Anderton}, {Andrews}, {Aubourg}, {Bailey}, {Bastien},
  {Bautista}, {Beers}, {Beifiori}, {Bender}, {Berlind}, {Beutler}, {Bhardwaj},
  {Bird}, {Bizyaev}, {Blake}, {Blanton}, {Blomqvist}, {Bochanski}, {Bolton},
  {Borde}, {Bovy}, {Shelden Bradley}, {Brand t}, {Brauer}, {Brinkmann},
  {Brownstein}, {Busca}, {Carithers}, {Carlberg}, {Carnero}, {Carr},
  {Chiappini}, {Chojnowski}, {Chuang}, {Comparat}, {Crepp}, {Cristiani},
  {Croft}, {Cuesta}, {Cunha}, {da Costa}, {Dawson}, {De Lee}, {Dean},
  {Delubac}, {Deshpande}, {Dhital}, {Ealet}, {Ebelke}, {Edmondson},
  {Eisenstein}, {Epstein}, {Escoffier}, {Esposito}, {Evans}, {Fabbian}, {Fan},
  {Favole}, {Femen{\'\i}a Castell{\'a}}, {Fern{\'a}ndez Alvar}, {Feuillet},
  {Filiz Ak}, {Finley}, {Fleming}, {Font-Ribera}, {Frinchaboy},
  {Galbraith-Frew}, {Garc{\'\i}a-Hern{\'a}ndez}, {Garc{\'\i}a P{\'e}rez}, {Ge},
  {G{\'e}nova-Santos}, {Gillespie}, {Girardi}, {Gonz{\'a}lez Hern{\'a}ndez},
  {Gott}, {Gunn}, {Guo}, {Halverson}, {Harding}, {Harris}, {Hasselquist},
  {Hawley}, {Hayden}, {Hearty}, {Herrero Dav{\'o}}, {Ho}, {Hogg}, {Holtzman},
  {Honscheid}, {Huehnerhoff}, {Ivans}, {Jackson}, {Jiang}, {Johnson},
  {Kinemuchi}, {Kirkby}, {Klaene}, {Kneib}, {Koesterke}, {Lan}, {Lang}, {Le
  Goff}, {Leauthaud}, {Lee}, {Lee}, {Long}, {Loomis}, {Lucatello}, {Lupton},
  {Ma}, {Mack}, {Mahadevan}, {Maia}, {Majewski}, {Malanushenko},
  {Malanushenko}, {Manchado}, {Manera}, {Maraston}, {Margala}, {Martell},
  {Masters}, {McBride}, {McGreer}, {McMahon}, {M{\'e}nard}, {M{\'e}sz{\'a}ros},
  {Miralda-Escud{\'e}}, {Miyatake}, {Montero-Dorta}, {Montesano}, {More},
  {Morrison}, {Muna}, {Munn}, {Myers}, {Nguyen}, {Nichol}, {Nidever},
  {Noterdaeme}, {Nuza}, {O'Connell}, {O'Connell}, {O'Connell}, {Olmstead},
  {Oravetz}, {Owen}, {Padmanabhan}, {Palanque-Delabrouille}, {Pan}, {Parejko},
  {Parihar}, {P{\^a}ris}, {Pepper}, {Percival}, {P{\'e}rez-R{\`a}fols}, {Dotto
  Perottoni}, {Petitjean}, {Pieri}, {Pinsonneault}, {Prada}, {Price-Whelan},
  {Raddick}, {Rahman}, {Rebolo}, {Reid}, {Richards}, {Riffel}, {Robin},
  {Rocha-Pinto}, {Rockosi}, {Roe}, {Ross}, {Ross}, {Rossi}, {Roy},
  {Rubi{\~n}o-Martin}, {Sabiu}, {S{\'a}nchez}, {Santiago}, {Sayres},
  {Schiavon}, {Schlegel}, {Schlesinger}, {Schmidt}, {Schneider}, {Schultheis},
  {Sellgren}, {Seo}, {Shen}, {Shetrone}, {Shu}, {Simmons}, {Skrutskie},
  {Slosar}, {Smith}, {Snedden}, {Sobeck}, {Sobreira}, {Stassun}, {Steinmetz},
  {Strauss}, {Streblyanska}, {Suzuki}, {Swanson}, {Terrien}, {Thakar},
  {Thomas}, {Thompson}, {Tinker}, {Tojeiro}, {Troup}, {Vandenberg}, {Vargas
  Maga{\~n}a}, {Viel}, {Vogt}, {Wake}, {Weaver}, {Weinberg}, {Weiner}, {White},
  {White}, {Wilson}, {Wisniewski}, {Wood-Vasey}, {Y{\`e}che}, {York}, {Zamora},
  {Zasowski}, {Zehavi}, {Zhao}, {Zheng}, \& {Zhu}}]{Ahn2014}
{Ahn}, C.~P., {Alexandroff}, R., {Allende Prieto}, C., {et~al.} 2014, \apjs,
  211, 17, \dodoi{10.1088/0067-0049/211/2/17}

\bibitem[{{Ahumada} {et~al.}(2020){Ahumada}, {Allende Prieto}, {Almeida},
  {Anders}, {Anderson}, {Andrews}, {Anguiano}, {Arcodia}, {Armengaud},
  {Aubert}, {Avila}, {Avila-Reese}, {Badenes}, {Balland }, {Barger},
  {Barrera-Ballesteros}, {Basu}, {Bautista}, {Beaton}, {Beers}, {Benavides},
  {Bender}, {Bernardi}, {Bershady}, {Beutler}, {Bidin}, {Bird}, {Bizyaev},
  {Blanc}, {Blanton}, {Boquien}, {Borissova}, {Bovy}, {Brand t}, {Brinkmann},
  {Brownstein}, {Bundy}, {Bureau}, {Burgasser}, {Burtin}, {Cano-D{\'\i}az},
  {Capasso}, {Cappellari}, {Carrera}, {Chabanier}, {Chaplin}, {Chapman},
  {Cherinka}, {Chiappini}, {Doohyun Choi}, {Chojnowski}, {Chung}, {Clerc},
  {Coffey}, {Comerford}, {Comparat}, {da Costa}, {Cousinou}, {Covey}, {Crane},
  {Cunha}, {da Silva Ilha}, {Dai}, {Damsted}, {Darling}, {Davidson}, {Davies},
  {Dawson}, {De}, {de la Macorra}, {De Lee}, {de Andrade Queiroz}, {Deconto
  Machado}, {de la Torre}, {Dell'Agli}, {du Mas des Bourboux},
  {Diamond-Stanic}, {Dillon}, {Donor}, {Drory}, {Duckworth}, {Dwelly},
  {Ebelke}, {Eftekharzadeh}, {Eigenbrot}, {Elsworth}, {Eracleous},
  {Erfanianfar}, {Escoffier}, {Fan}, {Farr}, {Fern{\'a}ndez-Trincado},
  {Feuillet}, {Finoguenov}, {Fofie}, {Fraser-McKelvie}, {Frinchaboy},
  {Fromenteau}, {Fu}, {Galbany}, {Garcia}, {Garc{\'\i}a-Hern{\'a}ndez}, {Garma
  Oehmichen}, {Ge}, {Geimba Maia}, {Geisler}, {Gelfand }, {Goddy},
  {Gonzalez-Perez}, {Grabowski}, {Green}, {Grier}, {Guo}, {Guy}, {Harding},
  {Hasselquist}, {Hawken}, {Hayes}, {Hearty}, {Hekker}, {Hogg}, {Holtzman},
  {Horta}, {Hou}, {Hsieh}, {Huber}, {Hunt}, {Ider Chitham}, {Imig}, {Jaber},
  {Jimenez Angel}, {Johnson}, {Jones}, {J{\"o}nsson}, {Jullo}, {Kim},
  {Kinemuchi}, {Kirkpatrick}, {Kite}, {Klaene}, {Kneib}, {Kollmeier}, {Kong},
  {Kounkel}, {Krishnarao}, {Lacerna}, {Lan}, {Lane}, {Law}, {Le Goff}, {Leung},
  {Lewis}, {Li}, {Lian}, {Lin}, {Long}, {Longa-Pe{\~n}a}, {Lundgren}, {Lyke},
  {Ted Mackereth}, {MacLeod}, {Majewski}, {Manchado}, {Maraston}, {Martini},
  {Masseron}, {Masters}, {Mathur}, {McDermid}, {Merloni}, {Merrifield},
  {M{\'e}sz{\'a}ros}, {Miglio}, {Minniti}, {Minsley}, {Miyaji}, {Mohammad},
  {Mosser}, {Mueller}, {Muna}, {Mu{\~n}oz-Guti{\'e}rrez}, {Myers}, {Nadathur},
  {Nair}, {Nandra}, {do Nascimento}, {Nevin}, {Newman}, {Nidever}, {Nitschelm},
  {Noterdaeme}, {O'Connell}, {Olmstead}, {Oravetz}, {Oravetz}, {Osorio},
  {Pace}, {Padilla}, {Palanque-Delabrouille}, {Palicio}, {Pan}, {Pan},
  {Parker}, {Paviot}, {Peirani}, {Pe{\~n}a Ram{\'r}ez}, {Penny}, {Percival},
  {Perez-Fournon}, {P{\'e}rez-R{\`a}fols}, {Petitjean}, {Pieri},
  {Pinsonneault}, {Poovelil}, {Povick}, {Prakash}, {Price-Whelan}, {Raddick},
  {Raichoor}, {Ray}, {Rembold}, {Rezaie}, {Riffel}, {Riffel}, {Rix}, {Robin},
  {Roman-Lopes}, {Rom{\'a}n-Z{\'u}{\~n}iga}, {Rose}, {Ross}, {Rossi}, {Rowland
  s}, {Rubin}, {Salvato}, {S{\'a}nchez}, {S{\'a}nchez-Menguiano},
  {S{\'a}nchez-Gallego}, {Sayres}, {Schaefer}, {Schiavon}, {Schimoia},
  {Schlafly}, {Schlegel}, {Schneider}, {Schultheis}, {Schwope}, {Seo},
  {Serenelli}, {Shafieloo}, {Shamsi}, {Shao}, {Shen}, {Shetrone}, {Shirley},
  {Silva Aguirre}, {Simon}, {Skrutskie}, {Slosar}, {Smethurst}, {Sobeck},
  {Sodi}, {Souto}, {Stark}, {Stassun}, {Steinmetz}, {Stello}, {Stermer},
  {Storchi-Bergmann}, {Streblyanska}, {Stringfellow}, {Stutz}, {Su{\'a}rez},
  {Sun}, {Taghizadeh-Popp}, {Talbot}, {Tayar}, {Thakar}, {Theriault}, {Thomas},
  {Thomas}, {Tinker}, {Tojeiro}, {Toledo}, {Tremonti}, {Troup}, {Tuttle},
  {Unda-Sanzana}, {Valentini}, {Vargas-Gonz{\'a}lez}, {Vargas-Maga{\~n}a},
  {V{\'a}zquez-Mata}, {Vivek}, {Wake}, {Wang}, {Weaver}, {Weijmans}, {Wild},
  {Wilson}, {Wilson}, {Wolthuis}, {Wood-Vasey}, {Yan}, {Yang}, {Y{\`e}che},
  {Zamora}, {Zarrouk}, {Zasowski}, {Zhang}, {Zhao}, {Zhao}, {Zheng}, {Zheng},
  {Zhu}, \& {Zou}}]{Ahumada2020}
{Ahumada}, R., {Allende Prieto}, C., {Almeida}, A., {et~al.} 2020, \apjs, 249,
  3, \dodoi{10.3847/1538-4365/ab929e}

\bibitem[{{Aparicio} \& {Gallart}(2004)}]{aparicio2004}
{Aparicio}, A., \& {Gallart}, C. 2004, \aj, 128, 1465, \dodoi{10.1086/382836}

\bibitem[{{Bell} {et~al.}(2019){Bell}, {Cioni}, {Wright}, {Rubele}, {Nidever},
  {Tatton}, {van Loon}, {Ivanov}, {Subramanian}, {Oliveira}, {de Grijs},
  {Pennock}, {Choi}, {Zaritsky}, {Olsen}, {Niederhofer}, {Choudhury},
  {Mart{\'\i}nez-Delgado}, \& {Mu{\~n}oz}}]{Bell2019}
{Bell}, C. P.~M., {Cioni}, M.-R.~L., {Wright}, A.~H., {et~al.} 2019, \mnras,
  489, 3200, \dodoi{10.1093/mnras/stz2325}

\bibitem[{{Bell} {et~al.}(2020){Bell}, {Cioni}, {Wright}, {Rubele}, {Nidever},
  {Tatton}, {van Loon}, {Zaritsky}, {Choi}, {Choudhury}, {Clementini}, {de
  Grijs}, {Ivanov}, {Majewski}, {Marconi}, {Mart{\'\i}nez-Delgado}, {Massana},
  {Mu{\~n}oz}, {Niederhofer}, {No{\"e}l}, {Oliveira}, {Olsen}, {Pennock},
  {Ripepi}, {Subramanian}, \& {Vivas}}]{Bell2020}
---. 2020, \mnras, 499, 993, \dodoi{10.1093/mnras/staa2786}

\bibitem[{{Belokurov} \& {Erkal}(2019)}]{Belokurov2019}
{Belokurov}, V.~A., \& {Erkal}, D. 2019, \mnras, 482, L9,
  \dodoi{10.1093/mnrasl/sly178}

\bibitem[{{Bernard} {et~al.}(2018){Bernard}, {Schultheis}, {Di Matteo}, {Hill},
  {Haywood}, \& {Calamida}}]{bernard2018MNRAS}
{Bernard}, E.~J., {Schultheis}, M., {Di Matteo}, P., {et~al.} 2018, \mnras,
  477, 3507, \dodoi{10.1093/mnras/sty902}

\bibitem[{{Bertelli} {et~al.}(1992){Bertelli}, {Mateo}, {Chiosi}, \&
  {Bressan}}]{Bertelli1992}
{Bertelli}, G., {Mateo}, M., {Chiosi}, C., \& {Bressan}, A. 1992, \apj, 388,
  400, \dodoi{10.1086/171163}

\bibitem[{{Besla} {et~al.}(2007){Besla}, {Kallivayalil}, {Hernquist},
  {Robertson}, {Cox}, {van der Marel}, \& {Alcock}}]{Besla2007}
{Besla}, G., {Kallivayalil}, N., {Hernquist}, L., {et~al.} 2007, \apj, 668,
  949, \dodoi{10.1086/521385}

\bibitem[{{Besla} {et~al.}(2012){Besla}, {Kallivayalil}, {Hernquist}, {van der
  Marel}, {Cox}, \& {Kere{\v s}}}]{Besla2012}
---. 2012, \mnras, 421, 2109, \dodoi{10.1111/j.1365-2966.2012.20466.x}

\bibitem[{{Besla} {et~al.}(2018){Besla}, {Patton}, {Stierwalt},
  {Rodriguez-Gomez}, {Patel}, {Kallivayalil}, {Johnson}, {Pearson}, {Privon},
  \& {Putman}}]{besla2018}
{Besla}, G., {Patton}, D.~R., {Stierwalt}, S., {et~al.} 2018, \mnras, 480,
  3376, \dodoi{10.1093/mnras/sty2041}

\bibitem[{{Bica} {et~al.}(2008){Bica}, {Bonatto}, {Dutra}, \&
  {Santos}}]{Bica2008}
{Bica}, E., {Bonatto}, C., {Dutra}, C.~M., \& {Santos}, J.~F.~C. 2008, \mnras,
  389, 678, \dodoi{10.1111/j.1365-2966.2008.13612.x}

\bibitem[{{Bica} {et~al.}(2019){Bica}, {Westera}, {Kerber}, {Dias}, {Maia},
  {Santos}, \& {Barbuy}}]{Bica2019}
{Bica}, E., {Westera}, P., {Kerber}, L. d.~O., {et~al.} 2019, arXiv e-prints,
  arXiv:1907.08642.
\newblock \doarXiv{1907.08642}

\bibitem[{{Bitsakis} {et~al.}(2017){Bitsakis}, {Bonfini},
  {Gonz{\'a}lez-L{\'o}pezlira}, {Ram{\'\i}rez-Siordia}, {Bruzual}, {Charlot},
  {Maravelias}, \& {Zaritsky}}]{Bitsakis2017}
{Bitsakis}, T., {Bonfini}, P., {Gonz{\'a}lez-L{\'o}pezlira}, R.~A., {et~al.}
  2017, \apj, 845, 56, \dodoi{10.3847/1538-4357/aa8090}

\bibitem[{{Bitsakis} {et~al.}(2018){Bitsakis}, {Gonz{\'a}lez-L{\'o}pezlira},
  {Bonfini}, {Bruzual}, {Maravelias}, {Zaritsky}, {Charlot}, \&
  {Ram{\'\i}rez-Siordia}}]{Bitsakis2018}
{Bitsakis}, T., {Gonz{\'a}lez-L{\'o}pezlira}, R.~A., {Bonfini}, P., {et~al.}
  2018, \apj, 853, 104, \dodoi{10.3847/1538-4357/aaa244}

\bibitem[{{Bothun} \& {Thompson}(1988)}]{BothunThompson1988}
{Bothun}, G.~D., \& {Thompson}, I.~B. 1988, \aj, 96, 877,
  \dodoi{10.1086/114854}

\bibitem[{Bressan {et~al.}(2012)Bressan, Marigo, Girardi, Salasnich, {Dal
  Cero}, Rubele, \& Nanni}]{Bressan2012}
Bressan, A., Marigo, P., Girardi, L., {et~al.} 2012, MNRAS, 427, 127,
  \dodoi{10.1111/j.1365-2966.2012.21948.x}

\bibitem[{{Brocato} {et~al.}(1996){Brocato}, {Castellani}, {Ferraro},
  {Piersimoni}, \& {Testa}}]{Brocato1996}
{Brocato}, E., {Castellani}, V., {Ferraro}, F.~R., {Piersimoni}, A.~M., \&
  {Testa}, V. 1996, \mnras, 282, 614, \dodoi{10.1093/mnras/282.2.614}

\bibitem[{Brown {et~al.}(2018)Brown, Vallenari, Prusti, \&
  de~Bruijne}]{GaiaDR2}
Brown, A. G.~A., Vallenari, A., Prusti, T., \& de~Bruijne, J. H.~J. 2018,
  Astronomy {\&} Astrophysics, \dodoi{10.1051/0004-6361/201833051}

\bibitem[{{Carrera} {et~al.}(2008){Carrera}, {Gallart}, {Hardy}, {Aparicio}, \&
  {Zinn}}]{Carrera2008}
{Carrera}, R., {Gallart}, C., {Hardy}, E., {Aparicio}, A., \& {Zinn}, R. 2008,
  \aj, 135, 836, \dodoi{10.1088/0004-6256/135/3/836}

\bibitem[{{Choi} {et~al.}(2018{\natexlab{a}}){Choi}, {Nidever}, {Olsen},
  {Besla}, {Blum}, {Zaritsky}, {Cioni}, {van der Marel}, {Bell}, {Johnson},
  {Vivas}, {Walker}, {de Boer}, {No{\"e}l}, {Monachesi}, {Gallart}, {Monelli},
  {Stringfellow}, {Massana}, {Martinez-Delgado}, \& {Mu{\~n}oz}}]{Choi:2018b}
{Choi}, Y., {Nidever}, D.~L., {Olsen}, K., {et~al.} 2018{\natexlab{a}}, \apj,
  869, 125, \dodoi{10.3847/1538-4357/aaed1f}

\bibitem[{{Choi} {et~al.}(2018{\natexlab{b}}){Choi}, {Nidever}, {Olsen},
  {Blum}, {Besla}, {Zaritsky}, {van der Marel}, {Bell}, {Gallart}, {Cioni},
  {Johnson}, {Vivas}, {Saha}, {de Boer}, {No{\"e}l}, {Monachesi}, {Massana},
  {Conn}, {Martinez-Delgado}, {Mu{\~n}oz}, \& {Stringfellow}}]{Choi:2018a}
---. 2018{\natexlab{b}}, \apj, 866, 90, \dodoi{10.3847/1538-4357/aae083}

\bibitem[{{Cignoni} {et~al.}(2013){Cignoni}, {Cole}, {Tosi}, {Gallagher},
  {Sabbi}, {Anderson}, {Grebel}, \& {Nota}}]{Cignoni2013}
{Cignoni}, M., {Cole}, A.~A., {Tosi}, M., {et~al.} 2013, \apj, 775, 83,
  \dodoi{10.1088/0004-637X/775/2/83}

\bibitem[{{Cignoni} \& {Tosi}(2010)}]{cignonitosi2010}
{Cignoni}, M., \& {Tosi}, M. 2010, Advances in Astronomy, 2010, 158568,
  \dodoi{10.1155/2010/158568}

\bibitem[{{Cioni} {et~al.}(2011){Cioni}, {Clementini}, {Girardi}, {Guand
  alini}, {Gullieuszik}, {Miszalski}, {Moretti}, {Ripepi}, {Rubele}, {Bagheri},
  {Bekki}, {Cross}, {de Blok}, {de Grijs}, {Emerson}, {Evans}, {Gibson},
  {Gonzales-Solares}, {Groenewegen}, {Irwin}, {Ivanov}, {Lewis}, {Marconi},
  {Marquette}, {Mastropietro}, {Moore}, {Napiwotzki}, {Naylor}, {Oliveira},
  {Read}, {Sutorius}, {van Loon}, {Wilkinson}, \& {Wood}}]{Cioni2011}
{Cioni}, M. R.~L., {Clementini}, G., {Girardi}, L., {et~al.} 2011, \aap, 527,
  A116, \dodoi{10.1051/0004-6361/201016137}

\bibitem[{{Cutri} \& {et al.}(2013)}]{Cutri2013}
{Cutri}, R.~M., \& {et al.} 2013, VizieR Online Data Catalog, II/328

\bibitem[{{Dark Energy Survey Collaboration} {et~al.}(2016){Dark Energy Survey
  Collaboration}, {Abbott}, {Abdalla}, {Aleksi{\'c}}, {Allam}, {Amara},
  {Bacon}, {Balbinot}, {Banerji}, {Bechtol}, {Benoit-L{\'e}vy}, {Bernstein},
  {Bertin}, {Blazek}, {Bonnett}, {Bridle}, {Brooks}, {Brunner}, {Buckley-Geer},
  {Burke}, {Caminha}, {Capozzi}, {Carlsen}, {Carnero-Rosell}, {Carollo},
  {Carrasco-Kind}, {Carretero}, {Castander}, {Clerkin}, {Collett}, {Conselice},
  {Crocce}, {Cunha}, {D'Andrea}, {da Costa}, {Davis}, {Desai}, {Diehl},
  {Dietrich}, {Dodelson}, {Doel}, {Drlica-Wagner}, {Estrada}, {Etherington},
  {Evrard}, {Fabbri}, {Finley}, {Flaugher}, {Foley}, {Fosalba}, {Frieman},
  {Garc{\'{\i}}a-Bellido}, {Gaztanaga}, {Gerdes}, {Giannantonio}, {Goldstein},
  {Gruen}, {Gruendl}, {Guarnieri}, {Gutierrez}, {Hartley}, {Honscheid}, {Jain},
  {James}, {Jeltema}, {Jouvel}, {Kessler}, {King}, {Kirk}, {Kron}, {Kuehn},
  {Kuropatkin}, {Lahav}, {Li}, {Lima}, {Lin}, {Maia}, {Makler}, {Manera},
  {Maraston}, {Marshall}, {Martini}, {McMahon}, {Melchior}, {Merson}, {Miller},
  {Miquel}, {Mohr}, {Morice-Atkinson}, {Naidoo}, {Neilsen}, {Nichol}, {Nord},
  {Ogando}, {Ostrovski}, {Palmese}, {Papadopoulos}, {Peiris}, {Peoples},
  {Percival}, {Plazas}, {Reed}, {Refregier}, {Romer}, {Roodman}, {Ross},
  {Rozo}, {Rykoff}, {Sadeh}, {Sako}, {S{\'a}nchez}, {Sanchez}, {Santiago},
  {Scarpine}, {Schubnell}, {Sevilla-Noarbe}, {Sheldon}, {Smith}, {Smith},
  {Soares-Santos}, {Sobreira}, {Soumagnac}, {Suchyta}, {Sullivan}, {Swanson},
  {Tarle}, {Thaler}, {Thomas}, {Thomas}, {Tucker}, {Vieira}, {Vikram},
  {Walker}, {Wechsler}, {Weller}, {Wester}, {Whiteway}, {Wilcox}, {Yanny},
  {Zhang}, \& {Zuntz}}]{DES}
{Dark Energy Survey Collaboration}, {Abbott}, T., {Abdalla}, F.~B., {et~al.}
  2016, \mnras, 460, 1270, \dodoi{10.1093/mnras/stw641}

\bibitem[{{de Vaucouleurs}(1955{\natexlab{a}})}]{devaucouleurs55}
{de Vaucouleurs}, G. 1955{\natexlab{a}}, \aj, 60, 126, \dodoi{10.1086/107173}

\bibitem[{{de Vaucouleurs}(1955{\natexlab{b}})}]{devaucouleurs55b}
---. 1955{\natexlab{b}}, \aj, 60, 219, \dodoi{10.1086/107218}

\bibitem[{{Elson} {et~al.}(1997){Elson}, {Gilmore}, \& {Santiago}}]{Elson1997}
{Elson}, R. A.~W., {Gilmore}, G.~F., \& {Santiago}, B.~X. 1997, \mnras, 289,
  157, \dodoi{10.1093/mnras/289.1.157}

\bibitem[{El Youssoufi {et~al.}(2019)El Youssoufi, Cioni, Bell, Rubele,
  Bekki, de Grijs, Girardi, Ivanov, Matijevic, Niederhofer, Oliveira, Ripepi,
  Subramanian, \& van Loon}]{elyoussoufi2019}
El Youssoufi, D., Cioni, M.-R.~L., Bell, C. P.~M., {et~al.} 2019, Monthly
  Notices of the Royal Astronomical Society, 490, 1076,
  \dodoi{10.1093/mnras/stz2400}

\bibitem[{{Erkal} {et~al.}(2019){Erkal}, {Belokurov}, {Laporte}, {Koposov},
  {Li}, {Grillmair}, {Kallivayalil}, {Price-Whelan}, {Evans}, {Hawkins},
  {Hendel}, {Mateu}, {Navarro}, {del Pino}, {Slater}, {Sohn}, \& {Orphan Aspen
  Treasury Collaboration}}]{Erkal2019}
{Erkal}, D., {Belokurov}, V., {Laporte}, C.~F.~P., {et~al.} 2019, \mnras, 487,
  2685, \dodoi{10.1093/mnras/stz1371}

\bibitem[{{Flaugher} {et~al.}(2015){Flaugher}, {Diehl}, {Honscheid}, {Abbott},
  {Alvarez}, {Angstadt}, {Annis}, {Antonik}, {Ballester}, {Beaufore},
  {Bernstein}, {Bernstein}, {Bigelow}, {Bonati}, {Boprie}, {Brooks},
  {Buckley-Geer}, {Campa}, {Cardiel-Sas}, {Castand er}, {Castilla}, {Cease},
  {Cela-Ruiz}, {Chappa}, {Chi}, {Cooper}, {da Costa}, {Dede}, {Derylo},
  {DePoy}, {de Vicente}, {Doel}, {Drlica-Wagner}, {Eiting}, {Elliott}, {Emes},
  {Estrada}, {Fausti Neto}, {Finley}, {Flores}, {Frieman}, {Gerdes},
  {Gladders}, {Gregory}, {Gutierrez}, {Hao}, {Holland}, {Holm}, {Huffman},
  {Jackson}, {James}, {Jonas}, {Karcher}, {Karliner}, {Kent}, {Kessler},
  {Kozlovsky}, {Kron}, {Kubik}, {Kuehn}, {Kuhlmann}, {Kuk}, {Lahav}, {Lathrop},
  {Lee}, {Levi}, {Lewis}, {Li}, {Mand richenko}, {Marshall}, {Martinez},
  {Merritt}, {Miquel}, {Mu{\~n}oz}, {Neilsen}, {Nichol}, {Nord}, {Ogando},
  {Olsen}, {Palaio}, {Patton}, {Peoples}, {Plazas}, {Rauch}, {Reil}, {Rheault},
  {Roe}, {Rogers}, {Roodman}, {Sanchez}, {Scarpine}, {Schindler}, {Schmidt},
  {Schmitt}, {Schubnell}, {Schultz}, {Schurter}, {Scott}, {Serrano}, {Shaw},
  {Smith}, {Soares-Santos}, {Stefanik}, {Stuermer}, {Suchyta}, {Sypniewski},
  {Tarle}, {Thaler}, {Tighe}, {Tran}, {Tucker}, {Walker}, {Wang}, {Watson},
  {Weaverdyck}, {Wester}, {Woods}, {Yanny}, \& {DES
  Collaboration}}]{Flaugher2015}
{Flaugher}, B., {Diehl}, H.~T., {Honscheid}, K., {et~al.} 2015, \aj, 150, 150,
  \dodoi{10.1088/0004-6256/150/5/150}

\bibitem[{{Frogel} \& {Blanco}(1983)}]{FrogelBlanco1983}
{Frogel}, J.~A., \& {Blanco}, V.~M. 1983, \apjl, 274, L57,
  \dodoi{10.1086/184150}

\bibitem[{{Gaia Collaboration} {et~al.}(2016){Gaia Collaboration}, {Prusti},
  {de Bruijne}, {Brown}, {Vallenari}, {Babusiaux}, {Bailer-Jones}, {Bastian},
  {Biermann}, {Evans}, \& et~al.}]{Gaia2016}
{Gaia Collaboration}, {Prusti}, T., {de Bruijne}, J.~H.~J., {et~al.} 2016,
  \aap, 595, A1, \dodoi{10.1051/0004-6361/201629272}

\bibitem[{{Gallart} {et~al.}(1999){Gallart}, {Freedman}, {Aparicio},
  {Bertelli}, \& {Chiosi}}]{gallart1999}
{Gallart}, C., {Freedman}, W.~L., {Aparicio}, A., {Bertelli}, G., \& {Chiosi},
  C. 1999, \aj, 118, 2245, \dodoi{10.1086/301078}

\bibitem[{{Gallart} {et~al.}(2004){Gallart}, {Stetson}, {Hardy}, {Pont}, \&
  {Zinn}}]{Gallart2004}
{Gallart}, C., {Stetson}, P.~B., {Hardy}, E., {Pont}, F., \& {Zinn}, R. 2004,
  \apjl, 614, L109, \dodoi{10.1086/425866}

\bibitem[{{Gallart} {et~al.}(2008){Gallart}, {Stetson}, {Meschin}, {Pont}, \&
  {Hardy}}]{Gallart2008}
{Gallart}, C., {Stetson}, P.~B., {Meschin}, I.~P., {Pont}, F., \& {Hardy}, E.
  2008, \apjl, 682, L89, \dodoi{10.1086/590552}

\bibitem[{{Gallart} {et~al.}(2005){Gallart}, {Zoccali}, \&
  {Aparicio}}]{gallart2005}
{Gallart}, C., {Zoccali}, M., \& {Aparicio}, A. 2005, \araa, 43, 387,
  \dodoi{10.1146/annurev.astro.43.072103.150608}

\bibitem[{{Gardiner} \& {Hatzidimitriou}(1992)}]{GardinerHatzidimitriou1992}
{Gardiner}, L.~T., \& {Hatzidimitriou}, D. 1992, \mnras, 257, 195,
  \dodoi{10.1093/mnras/257.2.195}

\bibitem[{{Gatto} {et~al.}(2020){Gatto}, {Ripepi}, {Bellazzini}, {Cignoni},
  {Cioni}, {Dall'Ora}, {Longo}, {Marconi}, {Schipani}, \& {Tosi}}]{Gatto2020}
{Gatto}, M., {Ripepi}, V., {Bellazzini}, M., {et~al.} 2020, \mnras,
  \dodoi{10.1093/mnras/staa3003}

\bibitem[{{Glatt} {et~al.}(2010){Glatt}, {Grebel}, \& {Koch}}]{Glatt2010}
{Glatt}, K., {Grebel}, E.~K., \& {Koch}, A. 2010, \aap, 517, A50,
  \dodoi{10.1051/0004-6361/201014187}

\bibitem[{{G{\'o}rski} {et~al.}(2005){G{\'o}rski}, {Hivon}, {Banday}, {Wand
  elt}, {Hansen}, {Reinecke}, \& {Bartelmann}}]{Gorski2005}
{G{\'o}rski}, K.~M., {Hivon}, E., {Banday}, A.~J., {et~al.} 2005, \apj, 622,
  759, \dodoi{10.1086/427976}

\bibitem[{{Hardy} {et~al.}(1984){Hardy}, {Buonanno}, {Corsi}, {Janes}, \&
  {Schommer}}]{Hardy1984LMC}
{Hardy}, E., {Buonanno}, R., {Corsi}, C.~E., {Janes}, K.~A., \& {Schommer},
  R.~A. 1984, \apj, 278, 592, \dodoi{10.1086/161826}

\bibitem[{{Harris} \& {Zaritsky}(2001)}]{harris_zaritsky2001}
{Harris}, J., \& {Zaritsky}, D. 2001, \apjs, 136, 25, \dodoi{10.1086/321792}

\bibitem[{{Harris} \& {Zaritsky}(2004)}]{Harris2004}
---. 2004, \aj, 127, 1531, \dodoi{10.1086/381953}

\bibitem[{{Harris} \& {Zaritsky}(2009)}]{harris_zaritsky2009}
---. 2009, \aj, 138, 1243, \dodoi{10.1088/0004-6256/138/5/1243}

\bibitem[{{Harris}(1997)}]{Harris1997}
{Harris}, W.~E. 1997, VizieR Online Data Catalog, VII/202

\bibitem[{{Hidalgo} {et~al.}(2011){Hidalgo}, {Aparicio}, {Skillman}, {Monelli},
  {Gallart}, {Cole}, {Dolphin}, {Weisz}, {Bernard}, {Cassisi}, {Mayer},
  {Stetson}, {Tolstoy}, \& {Ferguson}}]{hidalgo2011}
{Hidalgo}, S.~L., {Aparicio}, A., {Skillman}, E., {et~al.} 2011, \apj, 730, 14,
  \dodoi{10.1088/0004-637X/730/1/14}

\bibitem[{{Hill} \& {Zaritsky}(2006)}]{hill_zaritsky2006}
{Hill}, A., \& {Zaritsky}, D. 2006, \aj, 131, 414, \dodoi{10.1086/498647}

\bibitem[{{Hodge}(1987)}]{Hodge1987LMC_field}
{Hodge}, P. 1987, \pasp, 99, 730, \dodoi{10.1086/132038}

\bibitem[{{Holtzman} {et~al.}(1999){Holtzman}, {Gallagher}, {Cole}, {Mould},
  {Grillmair}, {Ballester}, {Burrows}, {Clarke}, {Crisp}, {Evans}, {Griffiths},
  {Hester}, {Hoessel}, {Scowen}, {Stapelfeldt}, {Trauger}, \&
  {Watson}}]{Holtzman1999}
{Holtzman}, J.~A., {Gallagher}, John~S., I., {Cole}, A.~A., {et~al.} 1999, \aj,
  118, 2262, \dodoi{10.1086/301097}

\bibitem[{Ibata {et~al.}(2017)Ibata, McConnachie, Cuillandre, Fantin, Haywood,
  Martin, Bergeron, Beckmann, Bernard, Bonifacio, Caffau, Carlberg, Côté,
  Cabanac, Chapman, Duc, Durret, Famaey, Fabbro, Gwyn, Hammer, Hill, Hudson,
  Lançon, Lewis, Malhan, Matteo, McCracken, Mei, Mellier, Navarro, Pires,
  Pritchet, Reylé, Richer, Robin, Jannsen, Sawicki, Scott, Scottez, Spekkens,
  Starkenburg, Thomas, \& Venn}]{ibata2017}
Ibata, R.~A., McConnachie, A., Cuillandre, J.-C., {et~al.} 2017, The
  Astrophysical Journal, 848, 129, \dodoi{10.3847/1538-4357/aa8562}

\bibitem[{{Irwin}(1991)}]{Irwin1991}
{Irwin}, M.~J. 1991, in IAU Symposium, Vol. 148, The Magellanic Clouds, ed.
  R.~{Haynes} \& D.~{Milne}, 453

\bibitem[{Ivezic {et~al.}(2008)Ivezic, Sesar, Juric, Bond, Dalcanton, Rockosi,
  Yanny, Newberg, Beers, Prieto, Wilhelm, Lee, Sivarani, Norris, Bailer-Jones,
  Fiorentin, Schlegel, Uomoto, Lupton, Knapp, Gunn, Covey, Smith, Miknaitis,
  Doi, Tanaka, Fukugita, Kent, Finkbeiner, Munn, Pier, Quinn, Hawley, Anderson,
  Kiuchi, Chen, Bushong, Sohi, Haggard, Kimball, Barentine, Brewington,
  Harvanek, Kleinman, Krzesinski, Long, Nitta, Snedden, Lee, Harris, Brinkmann,
  Schneider, \& York}]{Ivezic2008}
Ivezic, Z., Sesar, B., Juric, M., {et~al.} 2008, ApJ, 684, 40.
\newblock \url{http://arxiv.org/abs/0804.3850}

\bibitem[{{Johnson} {et~al.}(2015){Johnson}, {Seth}, {Dalcanton}, {Wallace},
  {Simpson}, {Lintott}, {Kapadia}, {Skillman}, {Caldwell}, {Fouesneau},
  {Weisz}, {Williams}, {Beerman}, {Gouliermis}, \& {Sarajedini}}]{Johnson2015}
{Johnson}, L.~C., {Seth}, A.~C., {Dalcanton}, J.~J., {et~al.} 2015, \apj, 802,
  127, \dodoi{10.1088/0004-637X/802/2/127}

\bibitem[{{Kallivayalil} {et~al.}(2006){Kallivayalil}, {van der Marel},
  {Alcock}, {Axelrod}, {Cook}, {Drake}, \& {Geha}}]{Kallivayalil2006}
{Kallivayalil}, N., {van der Marel}, R.~P., {Alcock}, C., {et~al.} 2006, \apj,
  638, 772, \dodoi{10.1086/498972}

\bibitem[{{Kallivayalil} {et~al.}(2013){Kallivayalil}, {van der Marel},
  {Besla}, {Anderson}, \& {Alcock}}]{Kallivayalil2013}
{Kallivayalil}, N., {van der Marel}, R.~P., {Besla}, G., {Anderson}, J., \&
  {Alcock}, C. 2013, \apj, 764, 161, \dodoi{10.1088/0004-637X/764/2/161}

\bibitem[{{Kroupa}(2001)}]{Kroupa2001}
{Kroupa}, P. 2001, \mnras, 322, 231, \dodoi{10.1046/j.1365-8711.2001.04022.x}

\bibitem[{Luo {et~al.}(2015)Luo, Zhao, Zhao, Deng, Liu, Jing, Wang, Zhang, Shi,
  Cui, \& et~al.}]{Luo2015}
Luo, A.-L., Zhao, Y.-H., Zhao, G., {et~al.} 2015, Research in Astronomy and
  Astrophysics, 15, 1095–1124, \dodoi{10.1088/1674-4527/15/8/002}

\bibitem[{{Majewski} {et~al.}(2009){Majewski}, {Nidever}, {Mu{\~n}oz},
  {Patterson}, {Kunkel}, \& {Carlin}}]{Majewski2009}
{Majewski}, S.~R., {Nidever}, D.~L., {Mu{\~n}oz}, R.~R., {et~al.} 2009, in IAU
  Symposium, Vol. 256, The Magellanic System: Stars, Gas, and Galaxies, ed.
  J.~T. {Van Loon} \& J.~M. {Oliveira}, 51--56,
  \dodoi{10.1017/S1743921308028251}

\bibitem[{{Marigo} {et~al.}(2008){Marigo}, {Girardi}, {Bressan}, {Groenewegen},
  {Silva}, \& {Granato}}]{Marigo2008}
{Marigo}, P., {Girardi}, L., {Bressan}, A., {et~al.} 2008, \aap, 482, 883,
  \dodoi{10.1051/0004-6361:20078467}

\bibitem[{{Marigo} {et~al.}(2017){Marigo}, {Girardi}, {Bressan}, {Rosenfield},
  {Aringer}, {Chen}, {Dussin}, {Nanni}, {Pastorelli}, {Rodrigues}, {Trabucchi},
  {Bladh}, {Dalcanton}, {Groenewegen}, {Montalb{\'a}n}, \& {Wood}}]{marigo2017}
---. 2017, \apj, 835, 77, \dodoi{10.3847/1538-4357/835/1/77}

\bibitem[{{Martin} {et~al.}(2015){Martin}, {Nidever}, {Besla}, {Olsen},
  {Walker}, {Vivas}, {Gruendl}, {Kaleida}, {Mu{\~n}oz}, {Blum}, {Saha}, {Conn},
  {Bell}, {Chu}, {Cioni}, {de Boer}, {Gallart}, {Jin}, {Kunder}, {Majewski},
  {Martinez-Delgado}, {Monachesi}, {Monelli}, {Monteagudo}, {No{\"e}l},
  {Olszewski}, {Stringfellow}, {van der Marel}, \& {Zaritsky}}]{Martin:2015}
{Martin}, N.~F., {Nidever}, D.~L., {Besla}, G., {et~al.} 2015, \apjl, 804, L5,
  \dodoi{10.1088/2041-8205/804/1/L5}

\bibitem[{{Martin} {et~al.}(2016){Martin}, {Jungbluth}, {Nidever}, {Bell},
  {Besla}, {Blum}, {Cioni}, {Conn}, {Kaleida}, {Gallart}, {Jin}, {Majewski},
  {Martinez-Delgado}, {Monachesi}, {Mu{\~n}oz}, {No{\"e}l}, {Olsen},
  {Stringfellow}, {van der Marel}, {Vivas}, {Walker}, \&
  {Zaritsky}}]{Martin:2016}
{Martin}, N.~F., {Jungbluth}, V., {Nidever}, D.~L., {et~al.} 2016, \apjl, 830,
  L10, \dodoi{10.3847/2041-8205/830/1/L10}

\bibitem[{{Mart{\'\i}nez-Delgado} {et~al.}(2019){Mart{\'\i}nez-Delgado},
  {Katherina Vivas}, {Grebel}, {Gallart}, {Pieres}, {Bell}, {Zivick},
  {Lemasle}, {Clifton Johnson}, {Carballo-Bello}, {No{\"e}l}, {Cioni}, {Choi},
  {Besla}, {Schmidt}, {Zaritsky}, {Gruendl}, {Seibert}, {Nidever},
  {Monteagudo}, {Monelli}, {Hubl}, {van der Marel}, {Ballesteros},
  {Stringfellow}, {Walker}, {Blum}, {Bell}, {Conn}, {Olsen}, {Martin}, {Chu},
  {Inno}, {Boer}, {Kallivayalil}, {De Leo}, {Beletsky}, {Neyer}, \&
  {Mu{\~n}oz}}]{Martinez-Delgado:2019}
{Mart{\'\i}nez-Delgado}, D., {Katherina Vivas}, A., {Grebel}, E.~K., {et~al.}
  2019, \aap, 631, A98, \dodoi{10.1051/0004-6361/201936021}

\bibitem[{{Massana} {et~al.}(2020){Massana}, {No{\"e}l}, {Nidever}, {Erkal},
  {de Boer}, {Choi}, {Majewski}, {Olsen}, {Monachesi}, {Gallart}, {van der
  Marel}, {Ruiz-Lara}, {Zaritsky}, {Martin}, {Mu{\~n}oz}, {Cioni}, {Bell},
  {Bell}, {Stringfellow}, {Belokurov}, {Monelli}, {Walker},
  {Mart{\'\i}nez-Delgado}, {Vivas}, \& {Conn}}]{Massana2020}
{Massana}, P., {No{\"e}l}, N. E.~D., {Nidever}, D.~L., {et~al.} 2020, \mnras,
  498, 1034, \dodoi{10.1093/mnras/staa2451}

\bibitem[{{Mateo} {et~al.}(1986){Mateo}, {Hodge}, \& {Schommer}}]{Mateo1986}
{Mateo}, M., {Hodge}, P., \& {Schommer}, R.~A. 1986, \apj, 311, 113,
  \dodoi{10.1086/164757}

\bibitem[{{Mau} {et~al.}(2020){Mau}, {Cerny}, {Pace}, {Choi}, {Drlica-Wagner},
  {Santana-Silva}, {Riley}, {Erkal}, {Stringfellow}, {Adam{\'o}w}, {Carlin},
  {Gruendl}, {Hernandez-Lang}, {Kuropatkin}, {Li}, {Mart{\'\i}nez-V{\'a}zquez},
  {Morganson}, {Mutlu-Pakdil}, {Neilsen}, {Nidever}, {Olsen}, {Sand},
  {Tollerud}, {Tucker}, {Yanny}, {Zenteno}, {Allam}, {Barkhouse}, {Bechtol},
  {Bell}, {Balaji}, {Crnojevi{\'c}}, {Esteves}, {Ferguson}, {Gallart},
  {Hughes}, {James}, {Jethwa}, {Johnson}, {Kuehn}, {Majewski}, {Mao},
  {Massana}, {McNanna}, {Monachesi}, {Nadler}, {No{\"e}l}, {Palmese},
  {Paz-Chinchon}, {Pieres}, {Sanchez}, {Shipp}, {Simon}, {Soares-Santos},
  {Tavangar}, {van der Marel}, {Vivas}, {Walker}, \& {Wechsler}}]{mau2020}
{Mau}, S., {Cerny}, W., {Pace}, A.~B., {et~al.} 2020, \apj, 890, 136,
  \dodoi{10.3847/1538-4357/ab6c67}

\bibitem[{{Meschin} {et~al.}(2014){Meschin}, {Gallart}, {Aparicio}, {Hidalgo},
  {Monelli}, {Stetson}, \& {Carrera}}]{meschin2014}
{Meschin}, I., {Gallart}, C., {Aparicio}, A., {et~al.} 2014, \mnras, 438, 1067,
  \dodoi{10.1093/mnras/stt2220}

\bibitem[{{Monachesi} {et~al.}(2012){Monachesi}, {Trager}, {Lauer}, {Hidalgo},
  {Freedman}, {Dressler}, {Grillmair}, \& {Mighell}}]{monachesi2012}
{Monachesi}, A., {Trager}, S.~C., {Lauer}, T.~R., {et~al.} 2012, \apj, 745, 97,
  \dodoi{10.1088/0004-637X/745/1/97}

\bibitem[{{Monelli} {et~al.}(2010){Monelli}, {Hidalgo}, {Stetson}, {Aparicio},
  {Gallart}, {Dolphin}, {Cole}, {Weisz}, {Skillman}, {Bernard}, {Mayer},
  {Navarro}, {Cassisi}, {Drozdovsky}, \& {Tolstoy}}]{monelli2010}
{Monelli}, M., {Hidalgo}, S.~L., {Stetson}, P.~B., {et~al.} 2010, \apj, 720,
  1225, \dodoi{10.1088/0004-637X/720/2/1225}

\bibitem[{{Monteagudo} {et~al.}(2018){Monteagudo}, {Gallart}, {Monelli},
  {Bernard}, \& {Stetson}}]{Monteagudo2018}
{Monteagudo}, L., {Gallart}, C., {Monelli}, M., {Bernard}, E.~J., \& {Stetson},
  P.~B. 2018, \mnras, 473, L16, \dodoi{10.1093/mnrasl/slx158}

\bibitem[{Nidever \& Dorta(2020)}]{PHOTRED}
Nidever, D., \& Dorta, A. 2020, dnidever/PHOTRED: SMASH DR2 Release,  Zenodo,
  \dodoi{10.5281/ZENODO.4291682}

\bibitem[{Nidever {et~al.}(2020)Nidever, Olsen, Choi, \& {SMASH
  Team}}]{SMASHRED}
Nidever, D., Olsen, K., Choi, Y., \& {SMASH Team}. 2020, dnidever/SMASHRED:
  SMASH DR2,  Zenodo, \dodoi{10.5281/ZENODO.4291646}

\bibitem[{{Nidever} {et~al.}(2008){Nidever}, {Majewski}, \& {Butler
  Burton}}]{Nidever2008}
{Nidever}, D.~L., {Majewski}, S.~R., \& {Butler Burton}, W. 2008, \apj, 679,
  432, \dodoi{10.1086/587042}

\bibitem[{{Nidever} {et~al.}(2010){Nidever}, {Majewski}, {Butler Burton}, \&
  {Nigra}}]{Nidever2010}
{Nidever}, D.~L., {Majewski}, S.~R., {Butler Burton}, W., \& {Nigra}, L. 2010,
  \apj, 723, 1618, \dodoi{10.1088/0004-637X/723/2/1618}

\bibitem[{{Nidever} {et~al.}(2011){Nidever}, {Majewski}, {Mu{\~n}oz}, {Beaton},
  {Patterson}, \& {Kunkel}}]{Nidever2011}
{Nidever}, D.~L., {Majewski}, S.~R., {Mu{\~n}oz}, R.~R., {et~al.} 2011, \apjl,
  733, L10, \dodoi{10.1088/2041-8205/733/1/L10}

\bibitem[{{Nidever} {et~al.}(2017){Nidever}, {Olsen}, {Walker}, {Vivas},
  {Blum}, {Kaleida}, {Choi}, {Conn}, {Gruendl}, {Bell}, {Besla}, {Mu{\~n}oz},
  {Gallart}, {Martin}, {Olszewski}, {Saha}, {Monachesi}, {Monelli}, {de Boer},
  {Johnson}, {Zaritsky}, {Stringfellow}, {van der Marel}, {Cioni}, {Jin},
  {Majewski}, {Martinez-Delgado}, {Monteagudo}, {No{\"e}l}, {Bernard},
  {Kunder}, {Chu}, {Bell}, {Santana}, {Frechem}, {Medina}, {Parkash},
  {Ser{\'o}n Navarrete}, \& {Hayes}}]{Nidever2017}
{Nidever}, D.~L., {Olsen}, K., {Walker}, A.~R., {et~al.} 2017, \aj, 154, 199,
  \dodoi{10.3847/1538-3881/aa8d1c}

\bibitem[{{Nidever} {et~al.}(2019){Nidever}, {Olsen}, {Choi}, {de Boer},
  {Blum}, {Bell}, {Zaritsky}, {Martin}, {Saha}, {Conn}, {Besla}, {van der
  Marel}, {No{\"e}l}, {Monachesi}, {Stringfellow}, {Massana}, {Cioni},
  {Gallart}, {Monelli}, {Martinez-Delgado}, {Mu{\~n}oz}, {Majewski}, {Vivas},
  {Walker}, {Kaleida}, \& {Chu}}]{Nidever2019a}
{Nidever}, D.~L., {Olsen}, K., {Choi}, Y., {et~al.} 2019, \apj, 874, 118,
  \dodoi{10.3847/1538-4357/aafaf7}

\bibitem[{{No{\"e}l} {et~al.}(2015){No{\"e}l}, {Conn}, {Read}, {Carrera},
  {Dolphin}, \& {Rix}}]{Noel2015}
{No{\"e}l}, N.~E.~D., {Conn}, B.~C., {Read}, J.~I., {et~al.} 2015, \mnras, 452,
  4222, \dodoi{10.1093/mnras/stv1614}

\bibitem[{{No{\"e}l} \& {Gallart}(2007)}]{Noel2007}
{No{\"e}l}, N. E.~D., \& {Gallart}, C. 2007, \apjl, 665, L23,
  \dodoi{10.1086/521223}

\bibitem[{{Olsen}(1999)}]{Olsen1999}
{Olsen}, K. A.~G. 1999, \aj, 117, 2244, \dodoi{10.1086/300854}

\bibitem[{{Olszewski} {et~al.}(1996){Olszewski}, {Suntzeff}, \&
  {Mateo}}]{Olszewski1996}
{Olszewski}, E.~W., {Suntzeff}, N.~B., \& {Mateo}, M. 1996, \araa, 34, 511,
  \dodoi{10.1146/annurev.astro.34.1.511}

\bibitem[{{Padmanabhan} {et~al.}(2008){Padmanabhan}, {Schlegel}, {Finkbeiner},
  {Barentine}, {Blanton}, {Brewington}, {Gunn}, {Harvanek}, {Hogg},
  {Ivezi{\'c}}, {Johnston}, {Kent}, {Kleinman}, {Knapp}, {Krzesinski}, {Long},
  {Neilsen}, {Nitta}, {Loomis}, {Lupton}, {Roweis}, {Snedden}, {Strauss}, \&
  {Tucker}}]{Padmanabhan2008}
{Padmanabhan}, N., {Schlegel}, D.~J., {Finkbeiner}, D.~P., {et~al.} 2008, \apj,
  674, 1217, \dodoi{10.1086/524677}

\bibitem[{{Piatti}(2018{\natexlab{a}})}]{Piatti18_LMC}
{Piatti}, A.~E. 2018{\natexlab{a}}, \mnras, 475, 2553,
  \dodoi{10.1093/mnras/stx3344}

\bibitem[{{Piatti}(2018{\natexlab{b}})}]{Piatti18_SMC}
---. 2018{\natexlab{b}}, \mnras, 478, 784, \dodoi{10.1093/mnras/sty1249}

\bibitem[{{Piatti} \& {Geisler}(2013)}]{Piatti2013}
{Piatti}, A.~E., \& {Geisler}, D. 2013, \aj, 145, 17,
  \dodoi{10.1088/0004-6256/145/1/17}

\bibitem[{{Pietrinferni} {et~al.}(2004){Pietrinferni}, {Cassisi}, {Salaris}, \&
  {Castelli}}]{pietrinferni2004}
{Pietrinferni}, A., {Cassisi}, S., {Salaris}, M., \& {Castelli}, F. 2004, \apj,
  612, 168, \dodoi{10.1086/422498}

\bibitem[{{Pietrzynski} {et~al.}(1999){Pietrzynski}, {Udalski}, {Kubiak},
  {Szymanski}, {Wozniak}, \& {Zebrun}}]{Pietrzynski1999}
{Pietrzynski}, G., {Udalski}, A., {Kubiak}, M., {et~al.} 1999, \actaa, 49, 521.
\newblock \doarXiv{astro-ph/9912187}

\bibitem[{{Rest} {et~al.}(2005){Rest}, {Stubbs}, {Becker}, {Miknaitis},
  {Miceli}, {Covarrubias}, {Hawley}, {Smith}, {Suntzeff}, {Olsen}, {Prieto},
  {Hiriart}, {Welch}, {Cook}, {Nikolaev}, {Huber}, {Prochtor}, {Clocchiatti},
  {Minniti}, {Garg}, {Challis}, {Keller}, \& {Schmidt}}]{Rest2005}
{Rest}, A., {Stubbs}, C., {Becker}, A.~C., {et~al.} 2005, \apj, 634, 1103,
  \dodoi{10.1086/497060}

\bibitem[{{Ripepi} {et~al.}(2014){Ripepi}, {Cignoni}, {Tosi}, {Marconi},
  {Musella}, {Grado}, {Limatola}, {Clementini}, {Brocato}, {Cantiello},
  {Capaccioli}, {Cappellaro}, {Cioni}, {Cusano}, {Dall'Ora}, {Gallagher},
  {Grebel}, {Nota}, {Palla}, {Romano}, {Raimondo}, {Sabbi}, {Getman},
  {Napolitano}, {Schipani}, \& {Zaggia}}]{Ripepi2014}
{Ripepi}, V., {Cignoni}, M., {Tosi}, M., {et~al.} 2014, \mnras, 442, 1897,
  \dodoi{10.1093/mnras/stu918}

\bibitem[{{Rubele} {et~al.}(2018){Rubele}, {Pastorelli}, {Girardi}, {Cioni},
  {Zaggia}, {Marigo}, {Bekki}, {Bressan}, {Clementini}, {de Grijs}, {Emerson},
  {Groenewegen}, {Ivanov}, {Muraveva}, {Nanni}, {Oliveira}, {Ripepi}, {Sun}, \&
  {van Loon}}]{rubele2018}
{Rubele}, S., {Pastorelli}, G., {Girardi}, L., {et~al.} 2018, \mnras, 478,
  5017, \dodoi{10.1093/mnras/sty1279}

\bibitem[{{Ruiz-Lara} {et~al.}(2020){Ruiz-Lara}, {Gallart}, {Monelli},
  {Nidever}, {Dorta}, {Choi}, {Olsen}, {Besla}, {Bernard}, {Cassisi},
  {Massana}, {No{\"e}l}, {P{\'e}rez}, {Rusakov}, {Cioni}, {Majewski}, {van der
  Marel}, {Mart{\'\i}nez-Delgado}, {Monachesi}, {Monteagudo}, {Mu{\~n}oz},
  {Stringfellow}, {Surot}, {Vivas}, {Walker}, \& {Zaritsky}}]{RuizLara2020b}
{Ruiz-Lara}, T., {Gallart}, C., {Monelli}, M., {et~al.} 2020, \aap, 639, L3,
  \dodoi{10.1051/0004-6361/202038392}

\bibitem[{{Saha} {et~al.}(2010){Saha}, {Olszewski}, {Brondel}, {Olsen},
  {Knezek}, {Harris}, {Smith}, {Subramaniam}, {Claver}, {Rest}, {Seitzer},
  {Cook}, {Minniti}, \& {Suntzeff}}]{Saha2010}
{Saha}, A., {Olszewski}, E.~W., {Brondel}, B., {et~al.} 2010, \aj, 140, 1719,
  \dodoi{10.1088/0004-6256/140/6/1719}

\bibitem[{{Schlegel} {et~al.}(1998){Schlegel}, {Finkbeiner}, \&
  {Davis}}]{Schlegel1998}
{Schlegel}, D.~J., {Finkbeiner}, D.~P., \& {Davis}, M. 1998, \apj, 500, 525,
  \dodoi{10.1086/305772}

\bibitem[{{Skrutskie} {et~al.}(2006){Skrutskie}, {Cutri}, {Stiening},
  {Weinberg}, {Schneider}, {Carpenter}, {Beichman}, {Capps}, {Chester},
  {Elias}, {Huchra}, {Liebert}, {Lonsdale}, {Monet}, {Price}, {Seitzer},
  {Jarrett}, {Kirkpatrick}, {Gizis}, {Howard}, {Evans}, {Fowler}, {Fullmer},
  {Hurt}, {Light}, {Kopan}, {Marsh}, {McCallon}, {Tam}, {Van Dyk}, \&
  {Wheelock}}]{Skrutskie2006}
{Skrutskie}, M.~F., {Cutri}, R.~M., {Stiening}, R., {et~al.} 2006, \aj, 131,
  1163, \dodoi{10.1086/498708}

\bibitem[{{Smecker-Hane} {et~al.}(2002){Smecker-Hane}, {Cole}, {Gallagher}, \&
  {Stetson}}]{Smecker-Hane2002}
{Smecker-Hane}, T.~A., {Cole}, A.~A., {Gallagher}, John~S., I., \& {Stetson},
  P.~B. 2002, \apj, 566, 239, \dodoi{10.1086/337985}

\bibitem[{{Stetson}(1994)}]{Stetson1994}
{Stetson}, P.~B. 1994, \pasp, 106, 250, \dodoi{10.1086/133378}

\bibitem[{{Stryker}(1984)}]{Stryker1984field}
{Stryker}, L.~L. 1984, \apjs, 55, 127, \dodoi{10.1086/190950}

\bibitem[{{Tolstoy} {et~al.}(2009){Tolstoy}, {Hill}, \& {Tosi}}]{tolstoy2009}
{Tolstoy}, E., {Hill}, V., \& {Tosi}, M. 2009, \araa, 47, 371,
  \dodoi{10.1146/annurev-astro-082708-101650}

\bibitem[{{Torrealba} {et~al.}(2018){Torrealba}, {Belokurov}, {Koposov},
  {Bechtol}, {Drlica-Wagner}, {Olsen}, {Vivas}, {Yanny}, {Jethwa}, {Walker},
  {Li}, {Allam}, {Conn}, {Gallart}, {Gruendl}, {James}, {Johnson}, {Kuehn},
  {Kuropatkin}, {Martin}, {Martinez-Delgado}, {Nidever}, {No{\"e}l}, {Simon},
  {Stringfellow}, \& {Tucker}}]{torrealba2018}
{Torrealba}, G., {Belokurov}, V., {Koposov}, S.~E., {et~al.} 2018, \mnras, 475,
  5085, \dodoi{10.1093/mnras/sty170}

\bibitem[{{Udalski} {et~al.}(1992){Udalski}, {Szymanski}, {Kaluzny}, {Kubiak},
  \& {Mateo}}]{Udalski1992}
{Udalski}, A., {Szymanski}, M., {Kaluzny}, J., {Kubiak}, M., \& {Mateo}, M.
  1992, \actaa, 42, 253

\bibitem[{{Valdes} {et~al.}(2014){Valdes}, {Gruendl}, \& {DES
  Project}}]{Valdes2014}
{Valdes}, F., {Gruendl}, R., \& {DES Project}. 2014, Astronomical Society of
  the Pacific Conference Series, Vol. 485, {The DECam Community Pipeline}, ed.
  N.~{Manset} \& P.~{Forshay}, 379

\bibitem[{{Vivas} {et~al.}(2016){Vivas}, {Olsen}, {Blum}, {Nidever}, {Walker},
  {Martin}, {Besla}, {Gallart}, {van der Marel}, {Majewski}, {Kaleida},
  {Mu{\~n}oz}, {Saha}, {Conn}, \& {Jin}}]{Vivas:2016}
{Vivas}, A.~K., {Olsen}, K., {Blum}, R., {et~al.} 2016, \aj, 151, 118,
  \dodoi{10.3847/0004-6256/151/5/118}

\bibitem[{{Weisz} {et~al.}(2013){Weisz}, {Dolphin}, {Skillman}, {Holtzman},
  {Dalcanton}, {Cole}, \& {Neary}}]{Weisz2013}
{Weisz}, D.~R., {Dolphin}, A.~E., {Skillman}, E.~D., {et~al.} 2013, \mnras,
  431, 364, \dodoi{10.1093/mnras/stt165}

\bibitem[{{Werchan} \& {Zaritsky}(2011)}]{werchan_zaritsky2011}
{Werchan}, F., \& {Zaritsky}, D. 2011, \aj, 142, 48,
  \dodoi{10.1088/0004-6256/142/2/48}

\bibitem[{{Westerlund}(1970)}]{Westerlund1970}
{Westerlund}, B.~E. 1970, Vistas in Astronomy, 12, 335,
  \dodoi{10.1016/0083-6656(70)90047-4}

\bibitem[{{Zaritsky} {et~al.}(2002){Zaritsky}, {Harris}, {Thompson}, {Grebel},
  \& {Massey}}]{Zaritsky2002}
{Zaritsky}, D., {Harris}, J., {Thompson}, I.~B., {Grebel}, E.~K., \& {Massey},
  P. 2002, \aj, 123, 855, \dodoi{10.1086/338437}

\end{thebibliography}
